\RequirePackage{ifpdf}
\documentclass[hyper]{JHEP3}

\pdfoutput=1
\usepackage{epsfig}
\usepackage{cite}
\usepackage{graphicx}
\usepackage{subfigure}
\usepackage{inputenc}
\usepackage{amsmath}

\newcommand{\beq}{\begin{equation}}
\newcommand{\eeq}{\end{equation}}
\newcommand{\beqn}{\begin{eqnarray}}
\newcommand{\eeqn}{\end{eqnarray}}
\def\df{{\rm d}}
\newcommand{\bn}{{\bar n}}

\def\bnslash{\bar n\!\!\!\slash}

\newcommand{\mcdot}{\!\cdot\!}
\newcommand{\eq}[1]{Eq.~\eqref{eq:#1}}
\newcommand{\eqs}[2]{Eqs.~\eqref{eq:#1} and \eqref{eq:#2}}
\def\a{\alpha}

\def\nn{\nonumber\\}
\def\pd{\partial}

\title{Standard Model Parton Distributions at Very High Energies}

\author{Christian W.~Bauer$^{a,b}$, Nicolas Ferland$^a$ and Bryan R.~Webber$^c$\\
   $^a$Ernest Orlando Lawrence Berkeley National Laboratory, University of California, Berkeley, CA 94720, USA\\
   $^b$Theoretical Physics Department, CERN, Geneva, Switzerland\\   
   $^c$University of Cambridge, Cavendish Laboratory, J.J.\ Thomson Avenue, Cambridge, UK\\
        E-mail: \email {cwbauer@lbl.gov}, \email{nferland@lbl.gov}, \email{webber@hep.phy.cam.ac.uk}
        }

\received{\today}               
\accepted{\today}               

\preprint{Cavendish-HEP-17/03\\CERN-TH-2017-067}

\abstract{We compute the leading-order evolution of parton
  distribution functions for all the Standard Model fermions and
  bosons up to energy scales far above the electroweak scale, where
  electroweak symmetry is restored.  Our results include the 52 PDFs
  of the unpolarized proton, evolving according to the SU(3), SU(2),
  U(1), mixed SU(2)$\times$U(1) and Yukawa interactions.  We illustrate
  the numerical effects on parton distributions at large energies, and show 
  that this can lead 
  to  important corrections to parton luminosities at a future
  100 TeV collider.
}

\begin{document} 

\section{Introduction}
\label{sec:intro}
Experiments at the Large Hadron Collider are now probing the structure
of matter at scales comparable with, and even beyond, the
characteristic scale of electroweak symmetry breaking.  So far, no
evidence has been found for a breakdown of the Standard Model (SM)
in particle collisions.
Indeed, there is a logical possibility that the SM remains a good
description of hard scattering processes up to scales far beyond
those of any conceivable particle colliders.  It is therefore of interest
to examine the features predicted by the SM for collider events well
above the electroweak scale.  For this purpose, Monte Carlo event
generators including all the SM interactions on an equal footing are
necessary.  Such generators would be useful for investigating the
limits of LHC searches, the potential of possible future colliders and
cosmic processes at ultrahigh energies. 

To construct a general-purpose SM event generator,\footnote{For a
 review of existing generators, see ref.~\cite{Buckley:2011ms}.} the three phases of
a hard collision, namely initial-state parton showering, parton-parton
collision and final-state showering, need to be simulated including
all SM particles and interactions.  For the initial-state showering,
parton distribution functions (PDFs) for all the SM fermions and
bosons need to be computed and tabulated beforehand, so that
showering can be generated backwards from the hard process, guided by
the scale dependence of the PDFs~\cite{Sjostrand:1985xi,Ellis:1991qj}.

Recently, a final-state parton shower including 
emissions from all interactions in the Standard Model was 
developed~\cite{Chen:2016wkt}, which illustrated the importance of 
electroweak splittings at high energies. For initial-state radiation the 
generalization of the
DGLAP~\cite{Gribov:1972ri,Dokshitzer:1977sg,Altarelli:1977zs}
evolution equations using all the Standard Model interactions has been 
worked out in~\cite{Ciafaloni:2005fm}, but so far no numerical implementation 
of these results has been published.  

As already mentioned, understanding the DGLAP evolution of PDFs using
all interactions of the SM 
 is a required first step in developing a complete initial state 
parton shower.  Moreover, it 
already allows us to study many new qualitative features of
very high-energy processes, such as lepton-initiated processes in
hadron collisions
and the polarization induced by electroweak PDF evolution.

The inclusion of QED corrections into parton distributions is a well
established procedure~\cite{Spiesberger:1994dm,Martin:2004dh,Roth:2004ti,
Ball:2013hta,Sadykov:2014aua,Carrazza:2015dea,Schmidt:2015zda,Manohar:2016nzj}.
However, above the electroweak scale around 100 GeV, the contributions
of other electroweak bosons become non-negligible and new effects
appear~\cite{Ciafaloni:1998xg,Ciafaloni:1999ub,Ciafaloni:2000df,Ciafaloni:2000gm,Ciafaloni:2000rp,Ciafaloni:2001vu,Ciafaloni:2001mu,Ciafaloni:2003xf,Ciafaloni:2005fm,Ciafaloni:2006qu,Ciafaloni:2008cr,Ciafaloni:2009mm,Ciafaloni:2010ti,Forte:2015cia,Mangano:2016jyj}.
PDFs of leptons, vector and scalar bosons are generated dynamically, and left-
and right-handed fermions evolve differently.  There are also
comparable effects in  the third generation of quarks due to their Yukawa
interactions.  Some effects of the SU(2) interaction are
double-logarithmically enhanced,  due to the non-singlet nature of the incoming
states.

The PDF evolution equations for the full Standard Model have been
presented in Ref.~\cite{Ciafaloni:2005fm}.  In the present paper we recast those
equations in a form suitable for event generation
and solve them
numerically for a given set of input distributions at the
electroweak scale.  The resulting PDF set extends through the region
of interest for future colliders and well beyond, so that we can study the
onset of the regime where all the SM interactions start to become
comparable.

Our solutions to the SM evolution equations are obtained in the approximation
of exact SU(3)$\times$SU(2)$\times$U(1) symmetry.  That is, we neglect
fermion and Higgs masses and the Higgs vacuum expectation value, the
effects of these being power-suppressed at high scales.  We
impose an infra-red cutoff $m_V$ on interactions that involve the
emission of an electroweak vector boson, $V=W^i$ for SU(2) or $B$
for U(1).  Leading-order evolution kernels and one-loop running
couplings are used.  All the electroweak PDFs
are generated dynamically from the QCD plus photon PDFs,
starting from a matching scale $q_0\sim m_V$.  In practice we take
 $q_0=m_V=100$ GeV. 
For the evolution of the photon, we decompose its PDF into
$W^3$, $B$ and mixed $B/W^3$ components at the input scale, evolve
these components, and reconstruct the photon PDF from them at higher
scales using the running SU(2) and U(1) couplings. For the top quark,
we set the PDF to zero below the top mass scale and then use the
leading-order massless evolution kernels, as for other fermions.
This treatment of the transition region around the electroweak scale
is clearly over-simplified but it should give a reliable indication of
the magnitude of electroweak effects at higher energies. 

The accuracy of our resulting PDFs is leading logarithmic, with
subleading logarithmic effects included where possible, but not in a
complete way. Contributions to the evolution from the U(1), SU(3) and
Yukawa interactions are therefore correct at the single logarithmic
level. However, as mentioned above, the SU(2) interactions give rise
to double logarithmic effects in the PDF evolution, such that single
logarithmic effects in SU(2) non-singlet quantities are not fully under control.
 
The organization of the paper is as follows. In Sec.~\ref{sec:SMevol}
we define the relevant parton distribution functions for unpolarized
proton beams and the general form of their evolution equations,
paying particular attention to the conservation of momentum in the
presence of the cutoff $m_V$ for vector boson emission.  After
specifying all the necessary splitting functions and running
couplings, we write the explicit evolution equations associated with
the five interactions: SU(3), U(1), SU(2), Yukawa and mixed
U(1)$\times$SU(2), for all the SM partons in a flavor basis. As usual 
for DGLAP evolution, we do not include 4 point interactions which are 
suppressed at high energies. 

For a numerical implementation, as described in Sec.~\ref{sec:implement},
the flavor basis is not convenient, as too many coupled equations are
involved.  Instead we use the basis of conserved quantum numbers
introduced in Ref.~\cite{Ciafaloni:2005fm}. As shown there, the
double-logarithmic evolution of SU(2) non-singlet PDFs can then be
factored out, which stabilizes and accelerates the solution of the
equations.  In this way we are able to evolve all the SM PDFs to
arbitrarily high scales with satisfactory speed and precision.
In practice we evolve up to $10^8$ GeV, where the approach to
asymptotic behavior is well established.

In Sec.~\ref{sec:results}, we present a selection of results that
illustrate the extent to which electroweak effects change the behavior of the 
various PDFs. In particular, we show changes in the PDFs of strongly interacting
particles relative to pure QCD
evolution, and show the size of the PDFs for electroweak gauge bosons relative 
to the gluon PDF. Finally, we present results of the associated changes in parton-parton luminosities at
a 100 TeV $pp$ collider. 
Our conclusions are presented in Sec.~\ref{sec:conc}. 

\section{The evolution of parton distributions in the full Standard Model}\label{sec:SMevol}

\subsection{Definition of the parton distribution functions}

The standard definition of an $x$-weighted parton distribution is given by the matrix element
of a bi-local operator, separated along the lightcone. For fermions, one finds the standard definition, but without spin averaging as we are separating the fermions into left- and right-handed. Thus, each fermion has only one possible spin determined by its helicity and the sign of its momentum
\begin{eqnarray}
 f_i(x,\mu) &=& x \int\!\!\frac{dy}{2\pi}\: e^{-i\, 2x\bn\cdot p\, y}
  \big\langle p \big| \,\bar\psi^{(i)}(y) \,{\bnslash}\,
  \psi^{(i)}(-y)\big| p \big\rangle \,,\\
f_{\bar i}(x,\mu) &=& x \int\!\!\frac{dy}{2\pi}\: e^{-i\, 2x\bn\cdot p\, y}
\big\langle p \big| \,\psi^{(i)}(y) \,{\bnslash}\,
\bar \psi^{(i)}(-y)\big| p \big\rangle \,,
\end{eqnarray}
where $\mu$ is the renormalization scale.
Since we have separate left- and right-handed PDFs, for each generation there are a total of 8 quark 
PDFs and 6 lepton PDFs to consider, giving a total of 42 fermion PDFs. 

Parton distributions functions of the vector bosons are given by
\begin{eqnarray}
 f_V(x,\mu) &=& \frac{2}{\bn\mcdot p} \int\!\!\frac{dy}{2\pi}\:
  e^{-i\, 2x\bn\cdot p\, y}\, \bn_\mu\bn^\nu
  \big\langle p \big| \,V^{\mu\lambda}(y) 
  V_{\lambda\nu}(-y)\big| p \big\rangle\Big|_{\mbox{\footnotesize spin avg.}}
  \,.
\end{eqnarray}
Since SU(3) is unbroken, we consider a single PDF to describe the gluon field. 
For the SU(2) $\otimes$ U(1) symmetry, on the other hand, one needs to take 
the symmetry breaking into account. For the $W^+$ and $W^-$ boson we simply
include separate PDFs for each of the two gauge bosons. For the $B$ and $W_3$, 
however, one needs to be more careful to take the mixed contributions
of these two bosons into account. Such contributions arise from the fact
that the left-handed fermions and Higgs carry both isospin and
hypercharge. This implies that besides $B$ and $W_3$ PDFs one needs to
include a mixed PDF, which is given
by\footnote{Note that our definition of the mixed PDF $f_{BW}$ is the
  sum of $BW_3$ and $W_3B$ contributions, and similarly for the mixed
PDF $f_{\gamma Z}$.}
\begin{eqnarray}
 f_{BW}(x) &=& \frac{2}{\bn\mcdot p} \int\!\!\frac{dy}{2\pi}\:
  e^{-i\, 2x\bn\cdot p\, y}\, \bn^\mu\bn_\nu
  \big\langle p \big| \,B_{\mu\lambda}(y) 
  W_3^{\lambda\nu}(-y)\big| p \big\rangle\Big|_{\mbox{\footnotesize spin avg.}} + \rm{h.c.}
  \,.
\end{eqnarray}
From these PDFs one can then construct the PDF for the photon, the
transversely-polarized $Z^0$ and their mixed state as a transformation
of the PDF for the $B$, the $W_3$ and their mixed state. Using $A =
c_W B + s_W W_3$ and $Z^0 = - s_W B + c_W W_3$ one finds
\beqn\label{eq:fgamz}
\left(
\begin{array}{c}
	f_{\gamma} \\
	f_Z \\
	f_{\gamma Z} \\
\end{array}
\right)
=\left(
\begin{array}{ccc}
	c_W^2 & s_W^2 & c_W s_W  \\
	s_W^2 & c_W^2 & -c_W s_W \\
	-2c_W s_W & 2c_W s_W & c_W^2-s_W^2 \\
\end{array}
\right)
\left(
\begin{array}{c}
	f_{B} \\
	f_{W_3} \\
	f_{BW} \\
\end{array}
\right) \,.
\eeqn
For the electroweak input at scale $\mu=q_0$ we have $f_\gamma\neq 0$ and
$f_Z=f_{\gamma Z}=0$, so the input conditions at that scale are
\beq\label{eq:fBWinput}
f_B = c_W^2f_\gamma\,,\;\;\;
f_{W_3} = s_W^2f_\gamma\,,\;\;\;
f_{BW} = 2c_W s_Wf_\gamma\,.
\eeq
After evolving these three unbroken PDFs to a higher scale $q$, the
physical photon and  $Z^0$ PDFs are reconstructed there using the
corresponding running values of $c_W$ and $s_W$.

Finally, one needs to include PDFs for the scalar bosons. One writes
\begin{eqnarray}
 f_H(x) &=& x \int\!\!\frac{dy}{2\pi}\:
  e^{-i\, 2x\bn\cdot p\, y}\, 
  \big\langle p \big| \,\Phi(y) 
  \Phi(-y)\big| p \big\rangle
  \,, \nn
\end{eqnarray}
and PDFs for each of the 4 Higgs fields $H^0$, $\bar H^0$, $H^+$ and
$H^-$ are included.  
The relationship to the 4 Higgs fields in the unbroken basis to the physical Higgs
and the longitudinal gauge bosons is as follows: 
The $H^\pm$ PDFs correspond to those of the
longitudinally polarized $W^\pm$. 
In the notation of
Ref.~\cite{Ciafaloni:2005fm}, the neutral Higgs fields are
\beq
H^0 = \frac{(h-iZ_L)}{\sqrt 2}\,,\qquad  \bar H^0 = \frac{(h+iZ_L)}{\sqrt 2}\,,
\eeq
where $h$ and $Z_L$ represent the Higgs and the longitudinal $Z^0$
fields, respectively.  The corresponding PDFs are
\beqn
f_{H^0} &=& \frac 12\left[f_{h}+f_{Z_L}+i\left(f_{hZ_L}-f_{Z_Lh}\right)\right]\,,\\
f_{\bar H^0} &=& \frac 12\left[f_{h}+f_{Z_L}-i\left(f_{hZ_L}-f_{Z_Lh}\right)\right]\,,
\eeqn
and one can also define the mixed PDFs
\beqn
f_{H^0 \bar H^0} &=& \frac 12\left[f_{h}-f_{Z_L}-i\left(f_{hZ_L}+f_{Z_Lh}\right)\right]\,,\\
f_{\bar H^0 H^0} &=& \frac 12\left[f_{h}-f_{Z_L}+i\left(f_{hZ_L}+f_{Z_Lh}\right)\right]
\,.\eeqn
Both of these mixed PDF carry non-zero hypercharge, such 
that they are not produced by the DGLAP evolution in the unbroken gauge
theory as considered in this paper\footnote{They are only produced through insertions of the 
Higgs vacuum}. Thus, one immediately finds
\beq
f_h-f_{Z_L} = f_{hZ_L}+f_{Z_Lh} = 0\,,
\eeq
and 
\beqn\label{eq:fhphi3}
f_h=f_{Z_L} = \frac 12 (f_{H^0}+f_{\bar H^0})\,,\qquad
f_{hZ_L} = -f_{Z_Lh} = -\frac i2 (f_{H^0}-f_{\bar H^0})\,.
\eeqn

In summary, there are a total of 52 parton distribution functions that
need to be considered.  Apart from the QCD quark and gluon
distributions and the electroweak PDFs \eqref{eq:fBWinput}, all the other
SM PDFs are set to zero at scale $q_0=m_V$ and evolve according to the
generalized DGLAP equations presented below.

\subsection{General evolution equations}
\label{sec:General_Intro}
We consider the $x$-weighted PDFs of parton species $i$
at momentum fraction $x$ and scale $q$, $f_i(x,q)$. In general they satisfy evolution equations of the
following forms:
\beqn
\label{eq:genevol}
q\frac{\pd}{\pd q} f_i(x, q) &=& \sum_I  \frac{\alpha_{I}(q)}{\pi} \left[  P^V_{i,I}(q) \, f_i(x, q) +  \sum_j  C_{ij,I} 
\int_x^{z_{\rm max}^{ij,I}(q)} \!\!\! \df z \, P^R_{ij, I}(z) f_j(x/z, q) \right] \nn
&\equiv& \sum_I \left[q\frac{\pd}{\pd q}  f_{i}(x, q)\right]_I
\,.
\eeqn
Here, the sum over $I$ goes over the different interactions in the
Standard Model
 and the notation $\left[q\,\pd / \pd q f_{i}(x, q)\right]_I$ implies that we only keep the terms proportional to the coupling $\alpha_I$ when taking the derivative\footnote{Note that  $\left[ \ldots \right]_I$ is only  introduced for notational convenience and should not be interpreted as setting all other couplings to zero. In particular, the PDFs appearing on the right-hand side of \eq{genevol} still depend on the value of all coupling constants}. For the rest of the section, we will show the evolution of each $f_i(x, q)$. We choose $I = 1, 2, 3$ for the pure ${\rm U}(1)$, ${\rm SU}(2)$ and ${\rm SU}(3)$ gauge interactions, $I =Y$ for Yukawa interactions, and 
$I = M$ for the mixed interaction proportional to
\beq
\alpha_M(q) = \sqrt{\alpha_1(q)\, \alpha_2(q)}
\,.\eeq
The first contribution, proportional to $P^V_{i,I}$, denotes the virtual contribution to the PDF evolution (the disappearance
of a flavor $i$), while the second contribution is the real contribution (the appearance of flavor $i$ due to the 
splitting of a flavor $j$). The maximum value of $z$ in the integration of the real contribution depends on the 
type of splitting and interaction, and we choose
\beq
\label{eq:zmax}
z_{\rm max}^{ij,I}(q) = \Big\{
\begin{array}{ll}
1 - \frac{m_V}{q} & {\rm for}\, I = 1, 2, \,{\rm and}\, i, j \notin V \,{\rm or}\, i, j \in V
\\
1 & {\rm otherwise}
\end{array}
\,,\eeq
that is, we apply an infrared cutoff $m_V$, of the order of the
electroweak scale, when a $B$ or $W$ boson is emitted.  This regulates
the divergence of the splitting function for those emissions as $z\to
1$.  Such a cutoff is mandatory for $I=2$ because there are PDF
contributions that are SU(2) non-singlets.  The evolution equations
for SU(3) are regular in the absence of a cutoff, as hadron PDFs are
color singlets.  Similarly for U(1), the unpolarized PDFs have zero
hypercharge,\footnote{Although there can be contributions with
  non-zero hypercharge for transversely polarized
  beams~\cite{Ciafaloni:2005fm}.}  but we include the same cutoff for
$I = 1$, since the $B$ and $W_3$ are mixed in the physical $Z$ and
$\gamma$ states.

Note that the precise choice of the cutoff is somewhat arbitrary, and as already mentioned, we choose $m_V = 100$ GeV in this paper. Changing this value changes our results by subleading logarithmic effect, at the same level as other effects not included.  However, given that the SU(2) evolution is double logarithmic, this implies that the ambiguity is single logarithmic for the SU(2) coupling. By matching our results to fixed order, one would account for these term at first order in $\alpha_2$. This is beyond the scope of this paper.

While the flavor basis chosen above is the most intuitive basis, the fact that all 52 PDFs are coupled to one another makes it quite difficult to solve the evolution equations. To decouple some of the equations, it helps to change the basis such that the ingredients have quantum numbers that are conserved in the Standard Model. Choosing the total isospin $\mathbf T$ and $\mathrm{CP}$  as the quantum numbers, the PDFs for each set of quantum numbers required are shown in Table~\ref{tab:T_CPStates}. 
\begin{table}[h!]
\begin{center}
\begin{tabular}{l|l}
$\{\mathbf T, \mathrm{CP}\}$ & fields\\\hline
$\{0,  +\}$ & $2 n_g\times q_R\,, n_g\times \ell_R\,, n_g\times q_L\,, n_g\times \ell_L\,, g\,, W\,, B\,, H$  \\
$\{0,  -\}$ & $2n_g\times q_R\,, n_g\times \ell_R\,, n_g\times q_L\,, n_g\times \ell_L\,, H$\\
$\{1,  +\}$ & $n_g\times q_L\,, n_g\times \ell_L\,, BW, H$ \\
$\{1,  -\}$ & $n_g\times q_L\,, n_g\times \ell_L\,, W, H$\\
$\{2, +\}$ & $W$ \\
\end{tabular}
\end{center}
\caption{\label{tab:T_CPStates}The 52 PDFs required for the SM evolution can written in a basis with definite conserved quantum numbers. $(5 n_g+4)$ PDFs contribute to the $\{0, +\}$ state, $(5 n_g+1)$ to the $\{0, -\}$, $(2 n_g+2)$ to each to the $\{1, +\}$ and $\{1, -\}$ and 1 to the $\{2, +\}$. }
\end{table}

Note that in general there can be additional mixed PDFs, which however
are zero in our initial conditions and which are not generated in the
evolution. In particular, there can be states mixing left-and
right-handed fermions, but they are not present in the initial
condition when only considering unpolarized beams because those states
are not Lorentz scalar. Thus, we can drop these states from our
evolution. 

The sum of momenta of all non-mixed PDFs in the particle basis is conserved, since it is the momentum of the proton. Momentum conservation applies independently for each interaction
\beq
\label{eq:MomentumConservation}
\sum _{i \neq \rm{BW}} \int_0^1\!\!dx \, \left[ q\frac{\pd}{\pd q} f_{i}(x, q)\right]_I =0 \,  {\rm for}\, I = 1, 2, 3, Y, M  \,.
\eeq
This is equivalent to the sum over all $\mathbf T=0$, $\mathrm{CP}=+$ PDFs in the
isospin and CP basis because only these states contribute
 to a sum over the PDFs in the particle
basis. For the other values of $\mathbf T$ and $\mathrm{CP}$, the PDFs correspond to
differences of PDFs in the particle basis. For example an
isospin 1 PDF is added in PDF of an up-type fermion, but
subtracted in the down-type PDF, thus it has no effect on the sum. 

Combining Eqs.~(\ref{eq:genevol}) and (\ref{eq:MomentumConservation}) gives
\beqn
0 &=& \sum_i P^V_{i} \, \int_0^1\!\!dx \, f_{i}(x,q) + \sum_{i,j} C_{ij,I} \int_0^1 \! \df x \, \int_x^{z_{\rm max}^{ij,I}(q)} \!\!\! \df z \, P^R_{ij,I}(z) \, f_{j}(x/z, q)
\nn
&= & \sum_i P^V_{i} \, \int_0^1\!\!dx \, f_{i}(x,q) + \sum_{i,j} C_{ij,I} \int_0^{z_{\rm max}^{ij,I}(q)} \!\!\! \df z \, P^R_{ij,I}(z)\int_0^z\!\!dx f_{j}(x/z, q)
\nn
& = & \sum_i P^V_{i} \, \langle f_{i}(q) \rangle + \sum_{i,j} C_{ij,I} \int_0^{z_{\rm max}^{ij,I}(q)} \!\!\! z \, \df z \, P^R_{ij,I}(z) \langle f_{j}(q) \rangle
\,,\eeqn
where we have defined the momentum averaged PDF
\beq
\langle f_{i}(q) \rangle \equiv \int_0^1\!\!dx \, f_{i}(x,q)\,.
\eeq
Solving the equation  for each of the $\langle f_{i}(q) \rangle$,
since all the input particle PDFs can be set independently, we get
\beqn
\label{eq:PVirtualDef}
P^V_{i,I}(q) &=& - \sum_{j} C_{ji,I} \int_0^{z_{\rm max}^{ji,I}(q)} \!\!\! z\,\df z\,P^R_{ji,I}(z)
\,.
\eeqn
Thus, momentum conservation determines the factor  $P^V_{i,I} $ for all non-mixed fields in the particle basis. 

Note that the result from momentum conservation agrees up to power
corrections with the more traditional definition of the virtual
corrections as loop insertions on the fields of the PDF.
Summing over possible loops, one has
\begin{align}\label{eq:PVirtualTrad}
\tilde P^V_{f_i,I}(q) &=-C_{ff,I} \int_0^{z_{\rm max}^{ff,G/Y}(q)} \df z\, P^R_{ff,I}(z)
\\
\tilde P^V_{V_i,I}(q) &= -\frac{C_{VV,I}}{2}\int_0^{z_{\rm max}^{VV,I}(q)} \df z\, P^R_{VV,I}(z) - \sum_{j\in f, h} C_{j,V_i,I}\int_0^1 \df z\, P^R_{jV,I}(z) 
\\
\tilde P^V_{HH,I}(q) &= -C_{HH,I}\int_0^{z_{\rm max}^{HH,I}(q)} \df z\, P^R_{HH,I}(z) - \sum_f C_{fH,I}\int_0^1 \df z\, P^R_{fH,I}(z)
\,,\end{align}
where
\beq
C_{ff,I}=\sum_{j}C_{f_jf_i,I}
\eeq
and similarly for $C_{VV,I}$ and $C_{HH,I}$.  The sums in
\eq{PVirtualTrad} extend over particles, and not their anti-particles. 
To see that \eqs{PVirtualDef}{PVirtualTrad} agree with each other, we will work it out explicitly for the virtual contribution to a fermion. One uses for the fermions that $P^R_{Vf,I}(z) = P^R_{ff,I}(1-z)$  and $C_{ff,I} = C_{Vf,I}$ to obtain the correct relation:
\beqn
P^V_{f,I}(q) &=& - C_{ff,I} \left[\int_0^{z_{\rm max}} \!\!\!\! z \,  \df z \, P^R_{ff,I}(z) +  \int_0^1 \!\!\!\! z \, \df z \, P^R_{Vf,I}(z) \right]
\nn
&=& - C_{ff,I} \left[\int_0^{z_{\rm max}} \!\!\!\! z \,  \df z \, P^R_{ff,I}(z) +  \int_0^1 \!\!\!\! (1-z) \, \df z \, P^R_{ff,I}(z) \right]
\nn
&=& - C_{ff,I} \left[\int_0^{z_{\rm max}} \!\!\!\!\df z \,
  P^R_{ff,I}(z) +  \int_{z_{\rm max}}^1 \!\!\!\! (1-z) \, \df z \, P^R_{ff,I}(z) \right]
\nn
&=&\tilde P^V_{f,I}(q) + \ldots
\,,\eeqn
where $\ldots$ denotes power corrections in $1 - z_{\rm max}$. The
argument is exactly the same for $P_{H,I}^V(q)$, while for
$P_{V,I}^V(q)$ one simply uses that $P^R_{VV,I}(z)$ and
$P^R_{fV,I}(z)$, and $P^R_{hV,I}(z)$ and $P^R_{fH,I}(z)$, are
symmetric in $z \leftrightarrow 1-z$ to write $\int \! z \, \df z =
\int \! \df z / 2$.   In our implementation of the evolution
equations, we use \eq{PVirtualDef}, to ensure exact momentum
conservation without explicit power corrections.

Since the mixed PDF $f_{BW}$ is a pure $\mathbf{T} = 1$ state, it does not contribute to the
momentum sum. This implies that one cannot derive its associated virtual contribution
from momentum conservation. However, using the traditional definition
in terms of loops, one sees that in this case the U(1) and SU(2) virtual
corrections each apply to only one of the two fields involved, and therefore
\begin{align}
\tilde P^V_{BW,1}(q) = \frac{1}{2} P^V_{B,1}(q)\,, \qquad \tilde P^V_{BW,2}(q) = \frac{1}{2} P^V_{W,2}(q)
\,,\end{align}
while the virtual contribution is zero for the other interactions. 

One can simplify the general evolution equations in \eq{genevol} by defining a full Sudakov factor
\beq
\label{eq:SudakovDef1}
\Delta_{i}(q) = \exp\left[\sum_I  \int_{q_0}^q \frac{\df q'}{q'} \frac{\alpha_I(q')}{\pi} P^V_{i,I}(q') \right]
\,,
\eeq
as well as a partial Sudakov factor for each interaction
\beq
\label{eq:SudakovDef2}
\Delta_{i,I}(q) = \exp\left[ \int_{q_0}^q \frac{\df q'}{q'} \frac{\alpha_I(q')}{\pi} P^V_{i,I}(q') \right]
\,,
\eeq
where $q_0$ is an arbitrary cutoff, which for convenience we set equal
to $m_V$. This allows us to write
\begin{equation}
\label{eq:genevol2}
\left[ \Delta_{i,I}(q) \, q\frac{\pd}{\pd q} \frac{ f_{i}(x, q)}{\Delta_{i,I}(q)}\right]_I = \frac{\alpha_{I}(q)}{\pi} \sum_j  C_{ij,I} P^R_{ij, I} \otimes f_j \,,
\end{equation}
where again the notation $[\ldots]_I$ implies that only terms from the interaction $I$ are kept. This gives
\begin{eqnarray}
\label{eq:genevol3}
\Delta_{i}(q) \, q\frac{\pd}{\pd q} \left[ \frac{ f_{i}(x, q)}{\Delta_{i}(q)}\right] &=& \sum_I \left[ \Delta_{i,I}(q) \, q\frac{\pd}{\pd q} \frac{ f_{i}(x, q)}{\Delta_{i,I}(q)}\right]_I
\nn 
&=&\sum_I \frac{\alpha_{I}(q)}{\pi} \sum_j  C_{ij,I} P^R_{ij, I} \otimes f_j \,,
\end{eqnarray}
where
\beq
P^R_{ij, I} \otimes f_j \equiv \int_x^{z_{\rm max}^{ij,I}(q)} \!\!\! \df z \, P^R_{ij, I}(z) f_j(x/z, q) 
\,.
\eeq

\subsection{Splitting functions}
The splitting functions depend only on the type of particles, which for the Standard Model are the spin 1/2 fermions, denoted by $f$, spin 1 gauge bosons, denoted by $V$, as well as spin 0 Higgs bosons, denoted by $H$. 

Denoting the three gauge interactions of the Standard Model
collectively by $I =G$, the splitting functions involving gauge bosons are given by
\beqn
P^R_{ff,G}(z) &=& \frac{1 + z^2}{1-z} \,, \\
P^R_{Vf,G}(z) &=& P_{ff,G}(1-z)\,,\\
P^R_{fV,G}(z) &=& \frac{1}{2} \left[ z^2+ (1-z)^2\right]\,,\\
P^R_{VV,G}(z) &=& 2\left[\frac z{1-z}+\frac{1-z}z+z(1-z)\right]\\
P^R_{HH,G}(z) &=& \frac{2z}{1-z}\,,\\
P^R_{VH,G}(z) &=& P^R_{HH,G}(1-z)\,,\\
P^R_{HV,G}(z) &=& z(1-z)\,.
\eeqn
The factor of $1/2$ in $P_{fV}$ has to be included since we are considering fermions with definite chirality.  For the Yukawa interaction ($Y$), one obtains
\beqn
P^R_{ff,Y}(z) &=& \frac{1-z}{2} \,, \\
P^R_{Hf,Y}(z) &=& P^R_{ff,Y}(1-z)\,,\\
P^R_{fH,Y}(z) &=& \frac{1}{2}\,.
\eeqn

\subsection{Running couplings}
\label{sec:coup}
The one-loop running of the gauge couplings $\a_I$ ($I=1,2,3$) is given by
\beq
\frac{2\pi}{\a_I(q_2)} = \frac{2\pi}{\a_I(q_1)}
+\beta_I\ln\frac{q_2}{q_1}
\,,
\eeq
where, for $n_g$ generations and $n_H$ Higgs doublets, 
\beqn\label{eq:SMrun}
\beta_1 &=& -\frac 13\rho_1 = -\frac{20}9 n_g-\frac 16 n_H=-\frac{41}6,\\
\beta_2 &=&\frac 23\left(11-\rho_{V2}\right)=\frac{22}3-\frac 43
n_g-\frac 16 n_H=\frac{19}6,\\
\beta_3 &=& 11-\rho_3 = 11-\frac 43 n_g=7\,.
\eeqn
At scale $M_Z=91.2$ GeV we take
\beq
\sin^2\theta_W = \frac{\a_1}{\a_1+\a_2}=0.23\,,\;\;\;
\a =  \a_2\sin^2\theta_W =\frac 1{128}\,,\;\;\;
\a_3 = 0.118\,,
\eeq
which gives
\beq\label{eq:alphas}
\a_1(M_Z)=0.0101\,\;\;\;
\a_2(M_Z)=0.0340\,\;\;\;
\a_3(M_Z)=0.118\,.
\eeq

We set all Yukawa couplings to zero, except for the top Yukawa
coupling $\a_Y=y_t^2/4\pi$. Its running receives significant
Yukawa and QCD contributions:
\beq
q\frac{\pd\a_Y}{\pd q} =
\frac{\a_Y}{2\pi}\left(\beta_Y\a_Y-\beta_S\a_3\right)\,,
\eeq
where $\beta_Y=9/2$ and $\beta_S=8$. The solution is
\beq
\frac 1{\a_Y(q_2)} =\frac\delta{\a_3(q_2)}-\left[\frac\delta{\a_3(q_1)}-\frac 1{\a_Y(q_1)}\right]
\left[\frac{\a_3(q_1)}{\a_3(q_2)}\right]^\gamma \,,
\eeq
where
\beqn
\gamma&=&\frac{\beta_S}{\beta_3}=\frac{24}{33-4n_g}=\frac 87\,,\\
\delta &=& \frac{\beta_Y}{\beta_S-\beta_3}=\frac{27}{8n_g-18}=\frac 92\,.
\eeqn
We take $m_t(m_t)=163$ GeV, which implies $\a_Y(m_t)=0.0349$, and
$\a_3(m_t)=0.109$.

\subsection{$I = 3$: SU(3) interactions}
\label{subsec:General_SU3}
We start by considering the well known case of SU(3) interactions. The
relevant degrees of freedom are the gluon, as well as left and
right-handed quarks. The coupling constants are (with $C_F = 4/3$, $C_A = 3$, $T_R = 1/2$)
\beq
\label{eq:SU3Couplings}
C_{qq,3} = C_{gq,3} = C_F\,, \qquad C_{qg,3} = T_R\,, \qquad C_{gg,3} = C_A
\,.
\eeq
This gives for the evolution of a quark or gluon\footnote{From now on
  we omit the arguments of functions for brevity.}
\beqn\label{eq:su3evol1}
\left[\Delta_{q,3} \, q\frac{\pd}{\pd q}\frac{ f_{q}}{\Delta_{q,3}}\right]_3  &=& \frac{\a_3}\pi \left[C_F P^R_{ff,G} \otimes f_q
+T_R P^R_{fV,G}\otimes f_g\right],\\
\left[\Delta_{g,3} \, q\frac{\pd}{\pd q}\frac{ f_{g}}{\Delta_{g,3}}\right]_3  &=& \frac{\a_3}\pi\left[C_A P^R_{VV,G}\otimes f_g+\sum_f C_F
P^R_{Vf,G}\otimes f_q\right]\,.
\label{eq:su3evol2}
\eeqn

The Sudakov factor can be obtained from \eq{PVirtualDef} using the coupling constants in \eq{SU3Couplings}. This gives
\begin{align}
\label{eq:SudakovDef3}
P^V_{q,3}(q) &= -C_F \int_0^1 \! z \, \df z \, \left[ P^R_{ff,G}(z) + P^R_{Vf,G}(z) \right]\,,
\\
P^V_{g,3}(q) &= -\int_0^1 \! z \, \df z \,  \left[ C_A \, P^R_{VV,G}(z) + 8 \, n_g \, T_R \, P^R_{fV,G}(z)\right] 
\,,
\end{align}
where we have used in the last line that there are 8 chiral quarks
plus antiquarks per generation. 

Since the gluon is massless, the upper limit in all the $z$
integrations is equal to 1 [see  \eq{zmax}]. This implies that the
convolutions $ P^R_{ff,G} \otimes f_q$ and $P^R_{VV,G}\otimes f_g$ in
\eqs{su3evol1}{su3evol2} are both divergent. However, at the same time
the virtual splitting functions that enter the Sudakov factors
$\Delta_{q,3}(q)$ and $\Delta_{g,3}(q)$ defined in \eq{SudakovDef2}
are also divergent, such that the divergences cancel in the evolution of the actual PDFs. Using +-distributions, as explained in Sec.~\ref{sec:implement}, one obtains evolution equations that are free of any divergences, and which can be implemented numerically. Alternatively, for parton shower implementation, one can impose a cutoff of the form \eq{zmax} with $m_V$ replaced by a small parameter $m_g>\Lambda_{\rm QCD}$.

\subsection{$I = 1$: U(1) interactions}
\label{subsec:General_U1}
For  ${\rm U}(1)$ the relevant degrees of freedom are left- and
right-handed fermions (denoted by the subscript $f$), as well as the
${\rm U}(1)$ gauge boson $B$. The couplings involving fermions and
gauge bosons are
\beq
C_{ff,1} = C_{Bf,1} = Y_f^2\,, \qquad C_{fB,1} = N_f \, Y_f^2\,, \qquad C_{BB,1} = 0
\eeq
where the hypercharges of the different fermions are given by
\beq
Y_{q_L} = \frac{1}{6}\,, \qquad Y_{u_R} = \frac{2}{3}\,, \qquad Y_{d_R} = -\frac{1}{3}\,, \qquad Y_{\ell_L} = -\frac{1}{2}\,, \qquad Y_{e_R} = -1
\,,
\eeq
and the color factor $N_f$ is equal to 3 for quarks and 1 for leptons. The couplings involving the Higgs bosons are
\begin{align}
C_{hh,1} = C_{Bh,1} = C_{hB,1} = \frac{1}{4}
\,,
\end{align}
where $h$ here stands for any of the four Higgs boson PDFs.

Plugging this into the general evolution equation gives
\beqn\label{eq:u1evol}
\left[\Delta_{f,1} \, q\frac{\pd}{\pd q}\frac{ f_{f}}{\Delta_{f,1}}\right]_1 &=& \frac{\a_1}\pi Y_i^2\left[P^R_{ff,G}\otimes f_f
+N_f P^R_{fV,G}\otimes f_B\right],\\
\left[\Delta_{B,1} \, q\frac{\pd}{\pd q}\frac{ f_{B}}{\Delta_{B,1}}\right]_1 &=& \frac{\a_1}\pi\left[\sum_f Y_f^2
P^R_{Vf,G}\otimes f_f + \frac{1}{4} \sum_h P^R_{VH,G} \otimes f_h \right]\,,\\
\left[\Delta_{H,1} \, q\frac{\pd}{\pd q}\frac{ f_{h}}{\Delta_{H,1}}\right]_1 &=& \frac{\a_1}\pi \frac{1}{4} \left[
P^R_{HH,G}\otimes f_h + P^R_{HV,G} \otimes f_B \right]\,.
\eeqn

The virtual splitting functions, required for the Sudakov factor are given by
\begin{align}
P^V_{f,1}(q) &= -Y_f^2 \left[  \int_0^{1-\frac{m_V}{q}} \! z \, \df z \, P^R_{ff,G}(z) + \int_0^1 \! z \, \df z \,  P^R_{Vf,G}(z) \right]\,,
\\
P^V_{B,1}(q) &= -n_g \left(\frac{11}{9}N_C+3 \right) \int_0^1 \! z \, \df z \,  P^R_{fV,G}(z) -  \int_0^1 \! z \, \df z \,  P^R_{HV,G}(z)\,,
\\
P^V_{H,1}(q) &= -  \frac{1}{4} \left[ \int_0^{1-\frac{m_V}{q}} \! z \, \df z \, P^R_{HH,G}(z) + \int_0^1 \! z \, \df z \,  P^R_{VH,G}(z) \right]
\,,\end{align}
where we have used in the second line that for each generation there are 4 left-handed quarks (one needs to count particles and antiparticles separately), 2 right-handed up-type quarks, 2 right-handed down-type quarks, 4 left-handed leptons and 2 right-handed electrons, and that there are a total of 4 Higgs bosons.

\subsection{$I = 2$: ${\rm SU}(2)$ interactions}
\label{subsec:General_SU2}
The SU(2) interactions are more complicated, since the emission of
$W^\pm$ bosons changes the flavor of the emitting particle. This,
combined with the SU(2) breaking in the input hadron PDFs, leads to
double-logarithmic scale dependence in the DGLAP evolution, rather than
only single-logarithmic dependence as in the evolution based on U(1) and SU(3). 

The relevant coupling constants are (where $u_L$ and $d_L$ denote any
up- and down-type left-handed fermion)
\beqn
C_{u_L d_L, 2} = C_{d_L u_L, 2} = C_{W^+ u_L, 2} = C_{W^- d_L, 2} &=& \frac 12\,,\\
C_{u_L u_L, 2} = C_{W_3 u_L, 2} = C_{d_L d_L, 2} = C_{W_3 d_L, 2}  &=& \frac 14\,,\\
C_{u_L W^+, 2}=C_{d_L W^-, 2} &=& N_f\frac 12\,,\\
C_{u_L W_3, 2} = C_{d_L W_3, 2} &=& N_f\frac 14\,,\\
C_{W^\pm W^\pm, 2} = C_{W^\pm W_3, 2} = C_{W_3 W^\pm, 2} &=& 1\,,
\eeqn
where as before the color factor $N_f=3$ for quarks, 1 for leptons.
The couplings of the $W_3$ state to the Higgs are given by
\beq
C_{hh,2} = C_{W_3 h,2} = C_{h W_3,2} = \frac{1}{4}\,,
\eeq
where again $h$ stands for any of the 4 Higgs bosons,
while those of the charged $W$ states are given by
\beqn
C_{H^+ H^0,2} &=& C_{H^0 H^+,2} = C_{H^+ W^+,2} = C_{W^+ H^+,2}\nn
&=& C_{H^0 W^-,2} = C_{W^- H^0,2} = \frac{1}{2}
\,.
\eeqn
The couplings for the charge-conjugate states are the same.

This gives for the evolution of the fermions
\beqn
\label{eq:SU2_fermion}
\left[\Delta_{f_L,2} \, q\frac{\pd}{\pd q}\frac{ f_{u_L}}{\Delta_{f_L,2}}\right]_2 &=&
\frac{\a_2}{\pi}\biggl\{
P^R_{ff,G} \otimes \left[\frac{f_{d_L}}{2}+\frac{f_{u_L}}{4}\right]
\nn
&& \qquad +N_f P_{fV,G}\otimes\left[\frac{f_{W^+}}{2}+\frac{f_{W_3}}{4}\right]\biggr\}\,,\\
\left[\Delta_{f_L,2} \, q\frac{\pd}{\pd q}\frac{ f_{d_L}}{\Delta_{f_L,2}}\right]_2 &=&
\frac{\a_2}{\pi}\biggl\{
P^R_{ff,G} \otimes \left[\frac{f_{u_L}}{2}+\frac{f_{d_L}}{4}\right]
\nn
&& \qquad
+N_f P_{fV,G}\otimes\left[\frac{f_{W^-}}{2}+\frac{f_{W_3}}{4}\right]\biggr\}
\,.\eeqn

For the $W^+$ and $W_3$ bosons we have
\beqn
\left[\Delta_{W,2} \, q\frac{\pd}{\pd q}\frac{ f_{W^+}}{\Delta_{W,2}}\right]_2 &=&
\frac{\a_2}{\pi}\biggl\{
P^R_{VV,G} \otimes \left[f_{W^+}+f_{W_3}\right] + \frac{1}{2} P^R_{VH,G} \otimes \left[ f_{H^+}+f_{\bar H^0}\right]
\nn
&& \qquad
+\sum_{\rm gen} \frac{1}{2} P_{Vf,G}\otimes\left[f_{u_L}+f_{\bar d_L}+f_{\nu_L}+f_{\bar\ell_L}\right]\biggr\}\,,\\
\left[\Delta_{W,2} \, q\frac{\pd}{\pd q}\frac{ f_{W_3}}{\Delta_{W,2}}\right]_2 &=&
\frac{\a_2}{\pi}\biggl\{
P^R_{VV,G} \otimes \left[f_{W^+}+f_{W^-}\right] + \frac{1}{4} P^R_{VH,G} \otimes \sum_h f_h
\nn
&& \qquad
+\frac{1}{4} \sum_{f_L} P^R_{Vf,G}\otimes f_{f_L}\biggr\}
\,,
\eeqn
where the sum in the last line is over all left-handed fermions and anti-fermions.
The equation for the $W^-$ can be obtained from that of the $W^+$ by charge conjugation.

Finally, for the Higgs bosons we have
\beqn
\left[\Delta_{H,2} \, q\frac{\pd}{\pd q}\frac{ f_{H^+}}{\Delta_{H,2}}\right]_2 &=&
\frac{\a_2}{\pi}\biggl\{
P^R_{HH,G} \otimes \left[\frac{f_{H^0}}{2} + \frac{f_{H^+}}{4}\right]
\nn
&& \qquad
+P_{HV,G}\otimes\left[\frac{f_{W^+}}{2}+\frac{f_{W_3}}{4}\right]\biggr\}\,,\\
\left[\Delta_{H,2} \, q\frac{\pd}{\pd q}\frac{ f_{H^0}}{\Delta_{H,2}}\right]_2 &=&
\frac{\a_2}{\pi}\biggl\{
P^R_{HH,G} \otimes \left[\frac{f_{H^+}}{2} + \frac{f_{H^0}}{4}\right]
\nn
&& \qquad
+P_{HV,G}\otimes\left[\frac{f_{W^-}}{2}+\frac{f_{W_3}}{4}\right]\biggr\}
\,.
\eeqn

The virtual splitting functions are
\begin{align}
\label{eq:SU2_virtual}
P^V_{f,2}(q) &= -\frac{3}{4} \left[  \int_0^{1-\frac{m_V}{q}} \! z \, \df z \, P^R_{ff,G}(z) + \int_0^{1} \! z \, \df z \, P^R_{Vf,G}(z) \right]\,,
\\
P^V_{W,2}(q) &= - 2  \int_0^{1-\frac{m_V}{q}} \! z \, \df z \,  P^R_{VV,G}(z) - n_g(N_C+1) \int_0^{1} \! z \, \df z \,  P^R_{fV,G}(z)  - \int_0^{1} \! z \, \df z \,  P^R_{HV,G}(z)\,,
\\
P^V_{H,2}(q) &= -  \frac{3}{4} \left[ \int_0^{1-\frac{m_V}{q}} \! z \, \df z \, P^R_{HH,G}(z) + \int_0^1 \! z \, \df z \,  P^R_{VH,G}(z) \right]
\,,\end{align}
from which the Sudakov factor can be constructed using \eq{SudakovDef2}. 

An important aspect of the SU(2) evolution equations is that, contrary to the other gauge groups, the dependence on the ratio $m_V / q$ does not cancel  between the real and virtual splitting functions. As an example, consider the evolution equation for an up-type fermion, given on the first line of \eq{SU2_fermion}, with the virtual contribution given by the first line of \eq{SU2_virtual}. The sum of the contributions of real and virtual splitting functions is given by
\beq
\frac{\a_2}{\pi}  \int_0^{1-\frac{m_V}{q}} \! \df z \, \frac{1}{4} P^R_{ff,G}(z)  \left[2 \, f_{d_L}(x/z)+f_{u_L}(x/z) - 3 \,f_{u_L}(x) \right]+\ldots
\,,\eeq
where $\ldots$ represents less singular terms.
Thus, the SU(2) breaking in the proton, which renders $f_u(z) \neq f_d(z)$, gives rise to a logarithmic dependence on $m_V/ q$, which leads to a double-logarithmic dependence upon integration over $q$. As we will see later, the effect of this dependence is to double-logarithmically suppress the SU(2) breaking effects at high energies. 

\subsection{$I = Y$: Yukawa interactions}
\label{subsec:General_Y}
The interaction of Higgs particles with fermions is described by the Yukawa interactions. In this work we only keep the top Yukawa coupling, setting all others to zero. This gives the following couplings
\beq
C_{q_L^3 t_R,Y} = C_{H^0 t_R, Y} = C_{H^+ t_R, Y} = C_{t_R q_L^3, Y} = C_{\bar H^0 t_L, Y} = C_{H^- b_L, Y} = 1
\,,
\eeq
where $q_L^3$ denotes either the left-handed top or bottom quark. We furthermore need
\beq
C_{t^R H^0, Y} = C_{t^R H^+, Y} = C_{t^L \bar H^0, Y} = C_{b_L H^-, Y} =  N_C
\,.
\eeq

This gives contributions to the top quark PDFs, as well as the left-handed bottom PDF:
\beqn
\left[\Delta_{q_L^3,Y} \, q\frac{\pd}{\pd q}\frac{ f_{t_L}}{\Delta_{q_L^3,Y}}\right]_Y &=&
\frac{\a_Y}{\pi}\biggl\{
P^R_{ff,Y} \otimes f_{t_R}  + N_C P_{fH,Y} \otimes f_{\bar H ^0} \biggr\}\,,\\
\left[\Delta_{t_R,Y} \, q\frac{\pd}{\pd q} \frac{ f_{t_R}}{\Delta_{t_R,Y}}\right]_Y &=&
\frac{\a_Y}{\pi}\biggl\{
P^R_{ff,Y} \otimes \left[f_{t_L} + f_{b_L} \right] + N_C P_{fH,Y} \otimes \left[f_{H ^0} + f_{H ^+} \right] \biggr\}\,,\\
\left[ \Delta_{q_L^3,Y} \, q\frac{\pd}{\pd q} \frac{ f_{b_L}}{\Delta_{q_L^3,Y}}\right]_Y &=&
\frac{\a_Y}{\pi}\biggl\{
P^R_{ff,Y} \otimes f_{t_R}  + N_C P_{fH,Y} \otimes f_{H ^-} \biggr\}\,.
\eeqn

It also contributes to the evolution of the Higgs bosons:
\beqn
\left[\Delta_{H,Y} \, q\frac{\pd}{\pd q}\frac{ f_{H^+}}{\Delta_{H,Y}}\right]_Y &=&
\frac{\a_Y}{\pi}
P^R_{Hf,Y} \otimes \left[ f_{t_R} + f_{\bar b_L} \right]\,,
\\
\left[\Delta_{H,Y} \, q\frac{\pd}{\pd q}\frac{ f_{H^0}}{\Delta_{H^0,Y}}\right]_Y &=&
\frac{\a_Y}{\pi}
P^R_{Hf,Y} \otimes \left[ f_{t_R} + f_{\bar t_L} \right] 
\,.
\eeqn

The Sudakov factors can be obtained using \eq{SudakovDef2} with
\begin{align}
P^V_{q_L^3,Y}(q) = \frac{1}{2} P^V_{t_R,Y}(q) &= -\int_0^{1} \! z \, \df z \, P^R_{ff,Y}(z) - \int_0^{1} \! z \, \df z \, P^R_{Hf,Y}(z)\,,
\\
P^V_{H,Y}(q) &= - 2 N_C  \int_0^{1} \! z \, \df z \,  P^R_{fH,Y}(z)
\,.\end{align}

\subsection{$I = M$: Mixed $B - W_3$ interactions}
\label{subsec:General_M}
Finally, we need to consider the evolution involving the mixed $BW$
boson PDF.  The non-vanishing couplings are
\beqn
C_{BWf_u,M} = -C_{BWf_d,M} &=& 2\frac{Y_f}{2}\,,\\
C_{f_uBW,M} = - C_{f_dBW,M} &=& N_f \frac{Y_f}{2}\,,
\eeqn
where $f_u$ and $f_d$ represent the  up- and down-type left-handed fermions
and anti-fermions of all generations. Since $Y_{\bar f}=-Y_f$ and $T_{3\bar f}=-T_{3f}$, the
couplings for fermions and anti-fermions are identical.
 The factor of 2 in the first line comes from our definition of
$f_{BW}$ as the sum of $BW$ and $WB$ contributions.
The diagonal coefficients $C_{f_uf_u,M}$ and $C_{f_df_d,M}$ are zero
because  there is no vector boson with both U(1) and SU(2)
interactions. For the same reason, there are no Sudakov factors
associated with the mixed interaction. The couplings involving the Higgs bosons are
 \beqn
C_{BW H^+,M} =  -C_{BW H^0,M} &=& \frac 12\,,\\
C_{H^+ BW,M} = -C_{H^0 BW,M} &=& \frac{1}{4}\,,
\eeqn
 where, as for the fermions, the same relations hold for the
charge-conjugate states.

Plugging these into the general evolution equation gives
\beqn\label{eq:mixedevol}
\left[q\frac{\pd}{\pd q}f_{f_u}\right]_M &=& \frac{\a_M}\pi \frac{Y_f}{2}N_f
P^R_{fV,G}\otimes f_{BW}\,,\\
\left[q\frac{\pd}{\pd q} f_{f_d}\right]_M &=& -\frac{\a_M}\pi \frac{Y_f}{2} N_f
P^R_{fV,G}\otimes f_{BW}\,,\\
\left[q\frac{\pd}{\pd q}f_{BW}\right]_M &=& \frac{\a_M}\pi
\big[
\sum_{f_u} Y_f P^R_{Vf,G}\otimes f_{f_u} - \sum_{f_d} Y_f P^R_{Vf,G}\otimes f_{f_d} \nn
&& \qquad + \frac{1}{2} \sum_{h_u} P^R_{VH,G} \otimes f_{h_u} - \frac{1}{2} \sum_{h_d} P^R_{VH,G} \otimes f_{h_d} 
\big]\,,\\
\left[q\frac{\pd}{\pd q} f_{h_u}\right]_M &=& \frac{\a_M}\pi \frac{1}{4} P^R_{HV,G} \otimes f_{BW}\,,\\
\left[q\frac{\pd}{\pd q} f_{h_d}\right]_M &=& -\frac{\a_M}\pi \frac{1}{4} P^R_{HV,G} \otimes f_{BW}\,.
\eeqn
As already discussed, the mixed gauge field PDF $f_{BW}$ has Sudakov factors associated
with the U(1) and SU(2) interactions, given by \eq{SudakovDef2}.
Since there is no corresponding real emission term in the evolution
equation for $f_{BW}$, it evolves double-logarithmically and is
suppressed at high scales relative to the unmixed PDFs.

\section{Implementation details}
\label{sec:implement}
Our treatment assumes that the SM PDFs at very high energies can be
obtained by smoothly matching the broken and unbroken symmetry regimes
at a matching scale $q_0\sim m_V$, which in practice we take to be
100 GeV.  Our input PDFs at 100 GeV are obtained as follows. We take
the CT14qed PDF set~\cite{Schmidt:2015zda} at 10 GeV and replace the
photon PDF by that of the LUXqed set~\cite{Manohar:2016nzj}.  We do not use
the CT14qed photon because the LUXqed photon, while being consistent with
CT14qed, has much smaller uncertanties and a smoother $x$ dependence.
The LUXqed PDF set combines the  PDF4LHC15\_nnlo\_100 parton
 set~\cite{Butterworth:2015oua}  with a determination of the photon PDF
from structure function and elastic form factor fits in electron-proton scattering.
However, we do not use the LUXqed partons, because being NNLO they are not
positive-definite, which we require for our LO treatment and is satisfied by CT14qed.

We evolve this hybrid CT14-LUX PDF set from 10 to 100 GeV using leading-order QCD
plus QED evolution, which incidentally generates the charged leptons.
The resulting parton, photon and lepton PDFs form our input to the
unbroken SM evolution upwards from 100 GeV.  The input left- and
right-handed fermion PDFs are identical. The input $W^3$, $B$ and
mixed $B/W^3$ PDFs are determined by the photon (and the absence of
the $Z^0$) at the matching scale according to \eq{fBWinput}.  The
remaining vector boson, neutrino and Higgs PDFs are all generated
dynamically starting from zero at the matching scale.

The equations given in Sections \ref{subsec:General_SU3} to
\ref{subsec:General_M} completely define the evolution of all parton
distribution functions in the unbroken symmetry regime. However, as
already explained, one
can rewrite the equations slightly to make them more amenable to a
numerical implementation. First, switching to a basis of states with
well-defined isospin decouples the set of 52 equations to some
degree. In this new basis another transformation eliminates
the double logarithmic sensitivity to the ratio $m_V / q$. Second, by
combining the virtual and real splitting functions into
+-distributions, one can reduce numerical sensitivity to the
cutoff of the $z$ integrations. We will now discuss these
simplifications in turn.

\subsection{Switching to a basis of conserved quantum numbers}
As we already explained in Section~\ref{sec:General_Intro}, the set of
52 evolution equations can be decoupled to some degree by switching to
a basis of well-defined isospin $\mathbf T$ and ${\mathrm{CP}}$.
Writing a fermion PDF with $\mathbf T$ and ${\mathrm{CP}}$ as
$f_i^{\mathbf T\mathrm{CP}}$, we write the left-handed fermions as
\begin{align}
\label{eq:fLIsospin}
f^{0+}_{f_L} &= \frac 14\left(f_{u_L}+f_{d_L}+f_{{\bar d}_L}+f_{{\bar u}_L}\right)\,,\qquad
     &f^{1+}_{f_L} &= \frac 14\left(f_{u_L}-f_{d_L}-f_{{\bar d}_L}+f_{{\bar u}_L}\right)\,,\\
f^{0-}_{f_L} &= \frac 14\left(f_{u_L}+f_{d_L}-f_{{\bar d}_L}-f_{{\bar u}_L}\right)\,,\qquad
     &f^{1-}_{f_L} &= \frac 14\left(f_{u_L}-f_{d_L}+f_{{\bar d}_L}-f_{{\bar u}_L}\right)
\,,
\end{align}
where $u_L$ and $d_L$ refer to left-handed up- and down-type fermions. Right-handed fermions are given by
\begin{align}
\label{eq:fRIsospin}
f^{0+}_{f_R} &= \frac 12\left(f_{f_R}+f_{{\bar f}_R}\right)\,,
     &f^{0-}_{f_R} &= \frac 12\left(f_{f_R}-f_{{\bar f}_R}\right)\,.
\end{align}
The SU(3) and U(1) boson PDFs have $\mathbf{T} = 0$, $\mathrm{CP}=+$
\begin{align}
f^{0+}_g &= f_g\,,
     &f^{0+}_B &= f_B\,,
\end{align}
while the SU(2) boson PDFs can have $\mathbf T = 0, 1, 2$ with respectively $\mathrm{CP} = +, -, +$
 \beqn
&&f^{0+}_W= \frac 13\left(f_{W^+}+f_{W^-}+f_{W^0}\right),\;\;\;
     f^{1-}_W = \frac 12\left(f_{W^+}-f_{W^-}\right),\\
&&f^{2+}_W = \frac 16\left(f_{W^+}+f_{W^-}-2f_{W^0}\right).
\eeqn
The mixed $BW$ boson state is a combination of $0^-$ and $1^-$ and
therefore its PDF has $\mathbf{T} = 1$, $\mathrm{CP}=+$
\beq
f^{1+}_{BW} = f_{BW}\,.
 \eeq
For the Higgs boson, one writes similarly to the fermions
 \beqn\label{eq:HIsospin}
&&f^{0+}_H = \frac 14\left(f_{H^+}+f_{H^0}+f_{\bar H^0}+f_{H^-}\right),\;\;\;
     f^{1+}_H= \frac 14\left(f_{H^+}-f_{H^0}-f_{\bar H^0}+f_{H^-}\right),\\
&&f^{0-}_H = \frac 14\left(f_{H^+}+f_{H^0}-f_{\bar H^0}-f_{H^-}\right),\;\;\;
     f^{1+}_H = \frac 14\left(f_{H^+}-f_{H^0}+f_{\bar
         H^0}-f_{H^-}\right)\,.
\eeqn
In terms of these states the longitudinal vector boson and
Higgs PDFs are then, using \eq{fhphi3},
\begin{align}
f_{W^+_L} &=f_H^{0+}+f_H^{1+}+f_H^{0-}+f_H^{1-}\,,\\
f_{W^-_L} &=f_H^{0+}+f_H^{1+}-f_H^{0-}-f_H^{1-}\,,\\
f_{Z_L}&=f_h=f_H^{0+}-f_H^{1+}\,.
\end{align}

\subsection{Cancellation of double-logarithmic dependence in evolution
  equations}
In the $\{\mathbf T,\mathrm{CP}\}$  basis the singular contributions to the evolution
equations (those that are proportional to the splitting functions
$P^R_{ff,G}(z)$,  $P^R_{VV,G}(z)$ and $P^R_{HH,G}(z)$, which diverge
in the limit $z \to 1$) are diagonal,
\begin{align}
\label{eq:divergentContribution}
\left[ \Delta_{i,I}\, q\frac{\pd}{\pd q} \frac{ f^{\mathbf T\mathrm{CP}}_{i}}{\Delta_{i,I}}\right]_I
& =\frac{\alpha_{I}}{\pi} D^{\mathbf T\mathrm{CP}}_{i,I} P^R_{ii, I} \otimes f^{\mathbf T\mathrm{CP}}_{i}  + \ldots\,,
\end{align}
such that the PDF multiplying the divergent splitting function is the
same as that appearing on the left-hand side. Here, as in
$f_i^{{\mathbf T} CP}$, the label $i$ now
refers to a parton species $f,V,H$ rather than a particular parton.
Recalling that the Sudakov factor takes the form
\beqn
\Delta_{i,I}(q)&=&\exp\left[ \int_{q_0}^q \frac{\df q'}{q'}
  \frac{\alpha_I(q')}{\pi} P^V_{i,I}(q') \right]\nn
&=&\exp\left[ -C_{i,I}\int_{q_0}^q \frac{\df q'}{q'}
  \frac{\alpha_I(q')}{\pi} \int_0^{z_{\rm max}^{ii,I}(q)} \!\!\!
  z\,\df z\,P^R_{ii,I}(z)+\ldots\right]\,,
\eeqn
where $\ldots$ represents less divergent terms, and
\beqn
C_{i,I} = \sum_{k \in i} C_{kl,I} \, {\rm for} \, l \in i \,,
\eeqn
where $k$ and $l$ are particular partons, 
we have
\beqn
\label{eq:divergentContribution2}
\left[q\frac{\pd}{\pd q} f^{\mathbf T\mathrm{CP}}_i\right]_I
&=&\frac{\alpha_{I}}{\pi} \left[D^{\mathbf T\mathrm{CP}}_{i,I} P^R_{ii, I}
\otimes f^{\mathbf T\mathrm{CP}}_{i}  +P^V_{i,I}f^{\mathbf T\mathrm{CP}}_{i}\right]+ \ldots\,,\nn
&=&\frac{\alpha_{I}}{\pi} \left[D^{\mathbf T\mathrm{CP}}_{i,I} P^+_{ii, I}
\otimes f^{\mathbf T\mathrm{CP}}_{i}  +\left(1-\frac{D^{\mathbf T\mathrm{CP}}_{i,I}}{C_{i,I}}\right)P^V_{i,I}f^{\mathbf T\mathrm{CP}}_{i}\right]+ \ldots\,,\nn
\eeqn
where
\beqn\label{eq:Pplusdef}
P^+_{ii, I} \otimes f_i&\equiv& P^R_{ii, I}\otimes f_i +\frac{P^V_{i,I}}{C_{i,I}}f_i\\
&=&\int_0^{z_{\rm max}^{ii,I}(q)}\!\!\! \df z\, \left[ P^R_{ii, I}(z) \theta(z > x) f(x/z, q) - z P^R_{ii, I}(z) f(x,q)\right]+\ldots\,.\nonumber
\eeqn
The +-prescription defined by \eq{Pplusdef} regulates the divergence
in the integrand as $z\to 1$ and therefore if we define the modifying factor
\beqn\label{eq:Fdef}
F^{\mathbf T\mathrm{CP}}_{i,I}(q) &=& \exp\left[ \left(1 -
    \frac{D^{\mathbf T\mathrm{CP}}_{i,I}}{C_{i,I}}\right)\int_{q_0}^q \frac{\df q'}{q'}
  \frac{\alpha_I(q')}{\pi} P^V_{i,I}(q')  \right]\nn
&=&\left[\Delta_{i,I}(q)\right]^{1- D^{\mathbf T\mathrm{CP}}_{i,I}/{C_{i,I}}}
\,,
\eeqn
then the evolution equation \eqref{eq:divergentContribution} becomes
\begin{align}
\label{eq:divergentContribution3}
\left[ F^{\mathbf T\mathrm{CP}}_{i,I}\, q\frac{\pd}{\pd q} \frac{ f^{\mathbf T\mathrm{CP}}_{i}}{F^{\mathbf T\mathrm{CP}}_{i,I}}\right]_I
& =\frac{\alpha_{I}}{\pi} D^{\mathbf T\mathrm{CP}}_{i,I} P^+_{ii, I} \otimes f^{\mathbf T\mathrm{CP}}_{i}  + \ldots\,,
\end{align}
with no logarithmic dependence  on $m_V / q$ on the right-hand side.

For all interactions except SU(2), one can show that $ D^{\mathbf
  T\mathrm{CP}}_{i,I} =C_{i,I}$, so that the modifying factor
\eqref{eq:Fdef} is unity\footnote{For the U(1) interaction one has $ D^{\mathbf
  T\mathrm{CP}}_{i,1} =C_{i,1} = 0$, and we choose to set the modifying factor to 1 in this case.}. For SU(2) we have explicitly:\footnote{Here we have used the numerical values for the
  Casimir operator eigenvalues for the corresponding SU(2)
  representations, $C_F^{\rm SU(2)}=3/4$,   $C_A^{\rm SU(2)}=2$.}
\beq
C_{f,2}=C_{H,2}=\frac 34\,,\;\;\; C_{V,2}=2\,,
\eeq
while
\begin{align}
D^{0\pm}_{f,2} &= D^{0\pm}_{H,2} = \frac 34\,,\;\;\; D^{1\pm}_{f,2} =
                 D^{1\pm}_{H,2} = -\frac 14\,,\\
D^{0+}_{V,2} &=2\,,\;\;\; D^{1-}_{V,2} =1\,,\;\;\; D^{2+}_{V,2} =-1\,,
\end{align}
so that
\begin{align}
F^{0\pm}_{f,2} &= F^{0\pm}_{H,2} =1\,,\;\;\; F^{1\pm}_{f/H,2} =\Delta_{f/H,2}^{4/3}\,,\\
F^{0+}_{V,2} &=1\,,\;\;\; F^{1-}_{V,2} =\Delta_{V,2}^{1/2}\,,\;\;\; F^{2+}_{V,2} =\Delta_{V,2}^{3/2}\,.
\end{align}
For the mixed PDF $f_{BW}$ we have $D^{1+}_{BW,2} =0$ and therefore
\beq
F^{1+}_{BW,2} =\Delta_{BW,2} =\Delta_{V,2}^{1/2}=F^{1-}_{V,2} 
\eeq

The equations finally used to evolve the PDFs in the conserved-quantum-number basis are given in
Appendix~\ref{app:forward}.

\section{Results}
\label{sec:results}
We begin by showing how the PDFs of strongly interacting particles are
changed by including the evolution of the full Standard
Model. Figure~\ref{fig:quarks} shows results on the evolution of
left- and right-handed quark PDFs, shown solid and dashed respectively,
normalized to their values assuming pure QCD evolution.
In each plot we show the results at three different scales, namely $q
= 10^4$ GeV, $q = 10^6$ GeV and $q = 10^8$ GeV. The values of
$10^6$ and $10^8$ GeV are of course far away from energy scales one can
reach at any collider in the near or distant future. However, showing
the results at such unattainable values helps to illustrate their
approach to asymptotic behavior.

All the light quarks (and antiquarks, not shown) evolve to lower
values compared to pure QCD at small $x$, due to an overall loss of
energy to the electroweak gauge bosons through the additional splittings
$q\to qW$ and $q\to qB$.  At higher $x$ values,
the up and down quarks (top row) exhibit different behaviors, with the
left-handed up PDF evolving more rapidly to lower values compared to pure QCD,
while the down quark eventually evolves to higher values.  This is because the
left-handed up and down distributions evolve towards each other, their
difference being double-logarithmically suppressed at high scales.
\FIGURE[h]{
 \centering
  \includegraphics[scale=0.45]{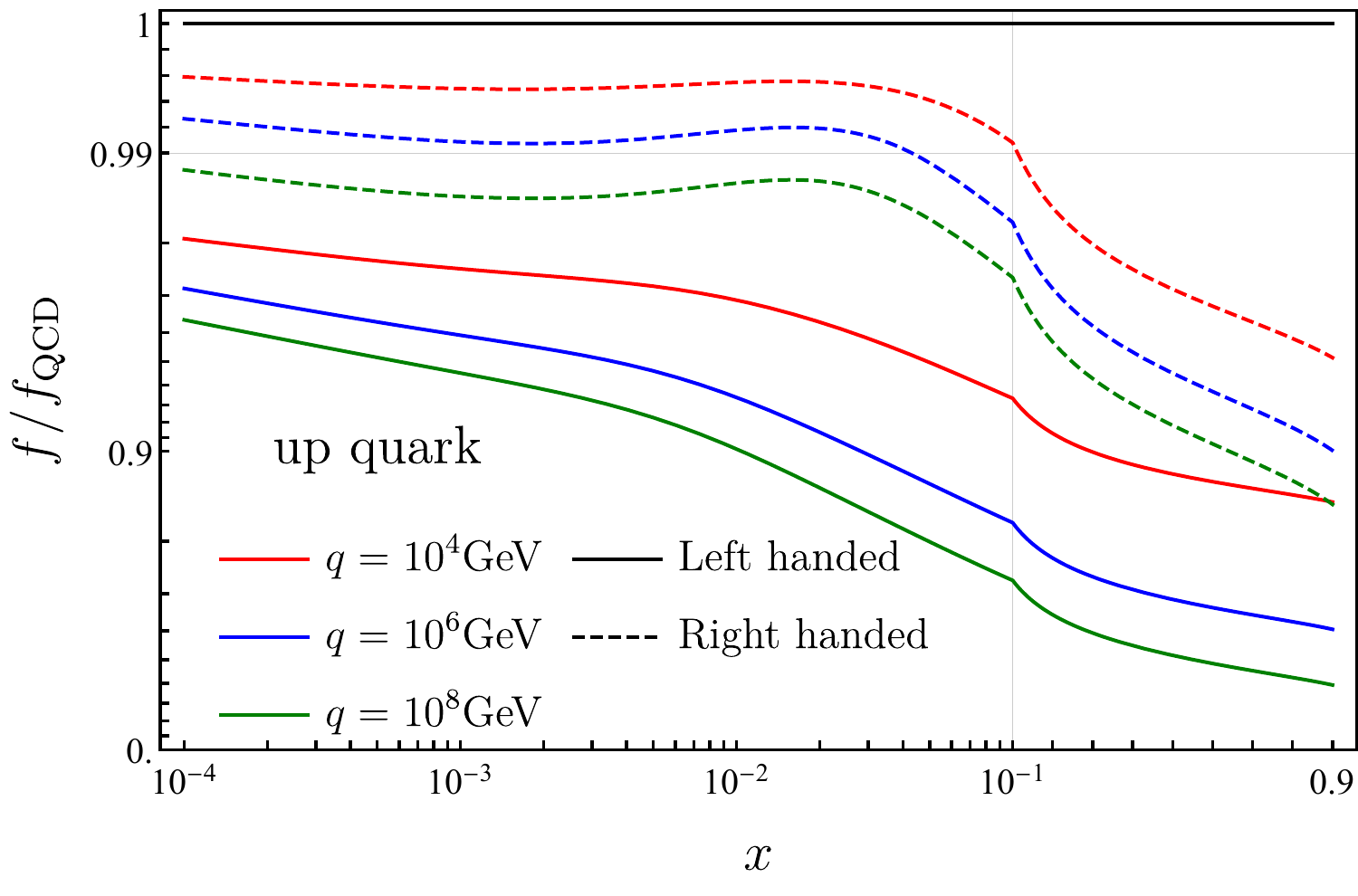}
  \includegraphics[scale=0.45]{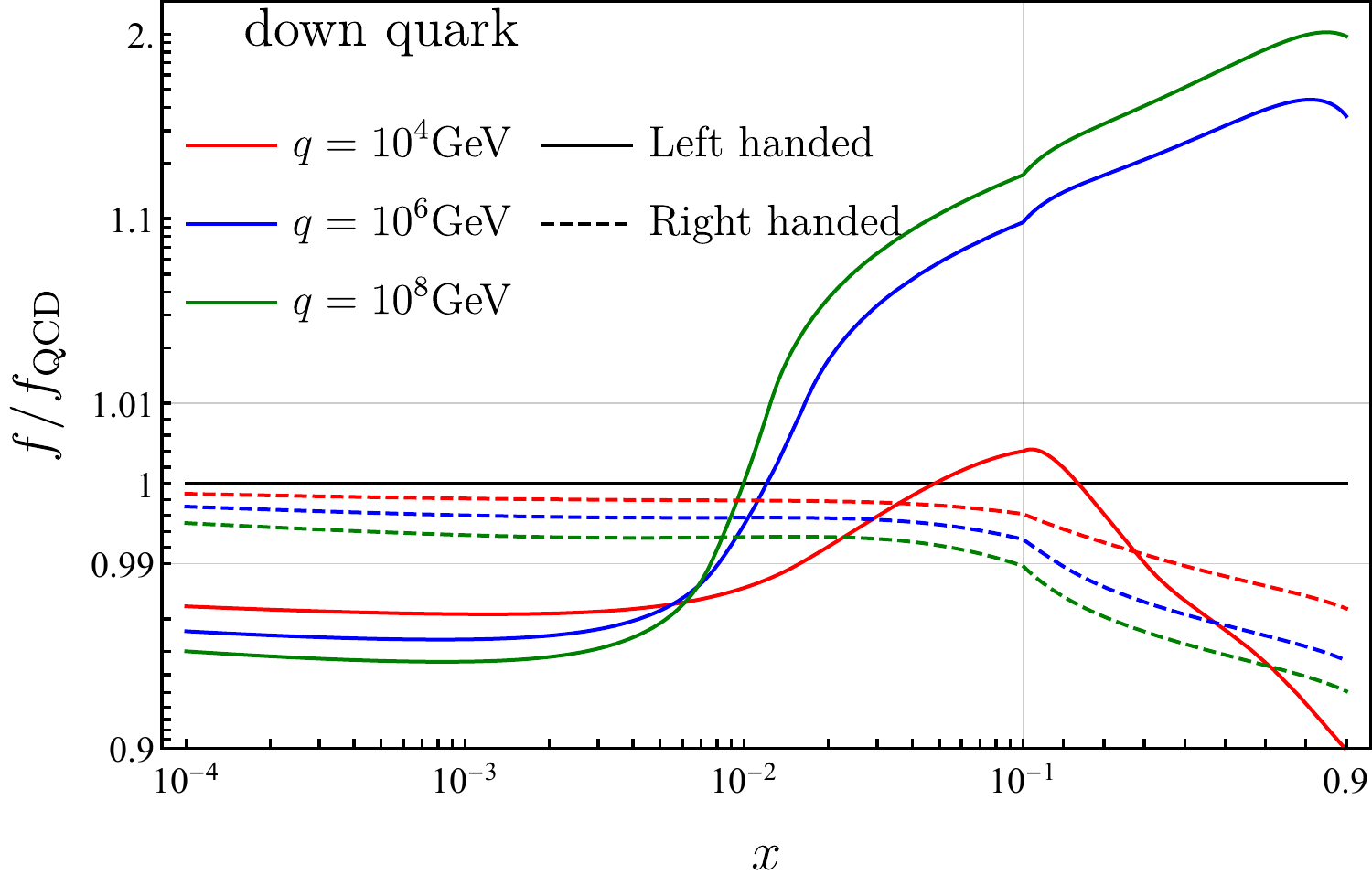}
  \includegraphics[scale=0.45]{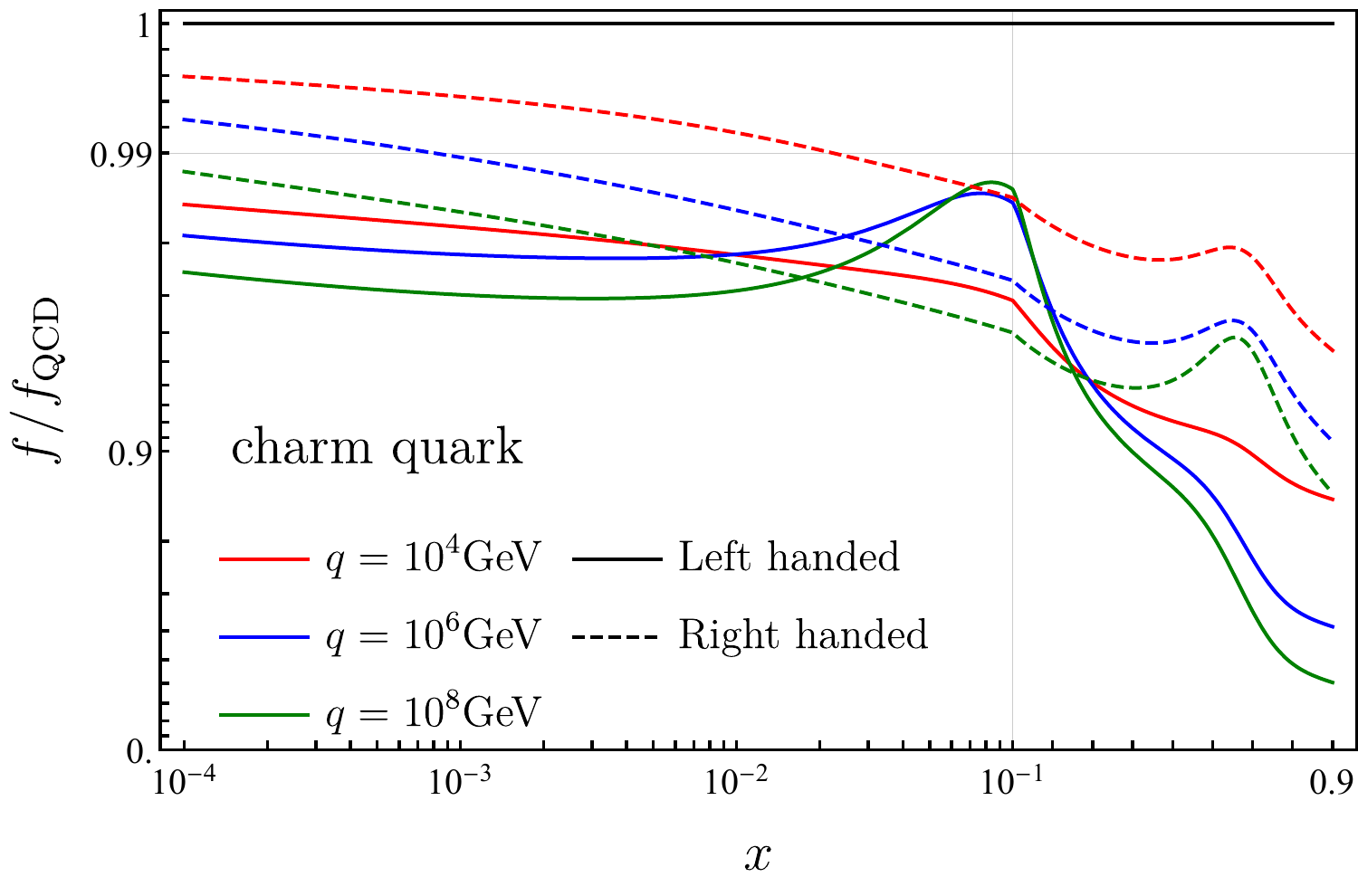}
  \includegraphics[scale=0.45]{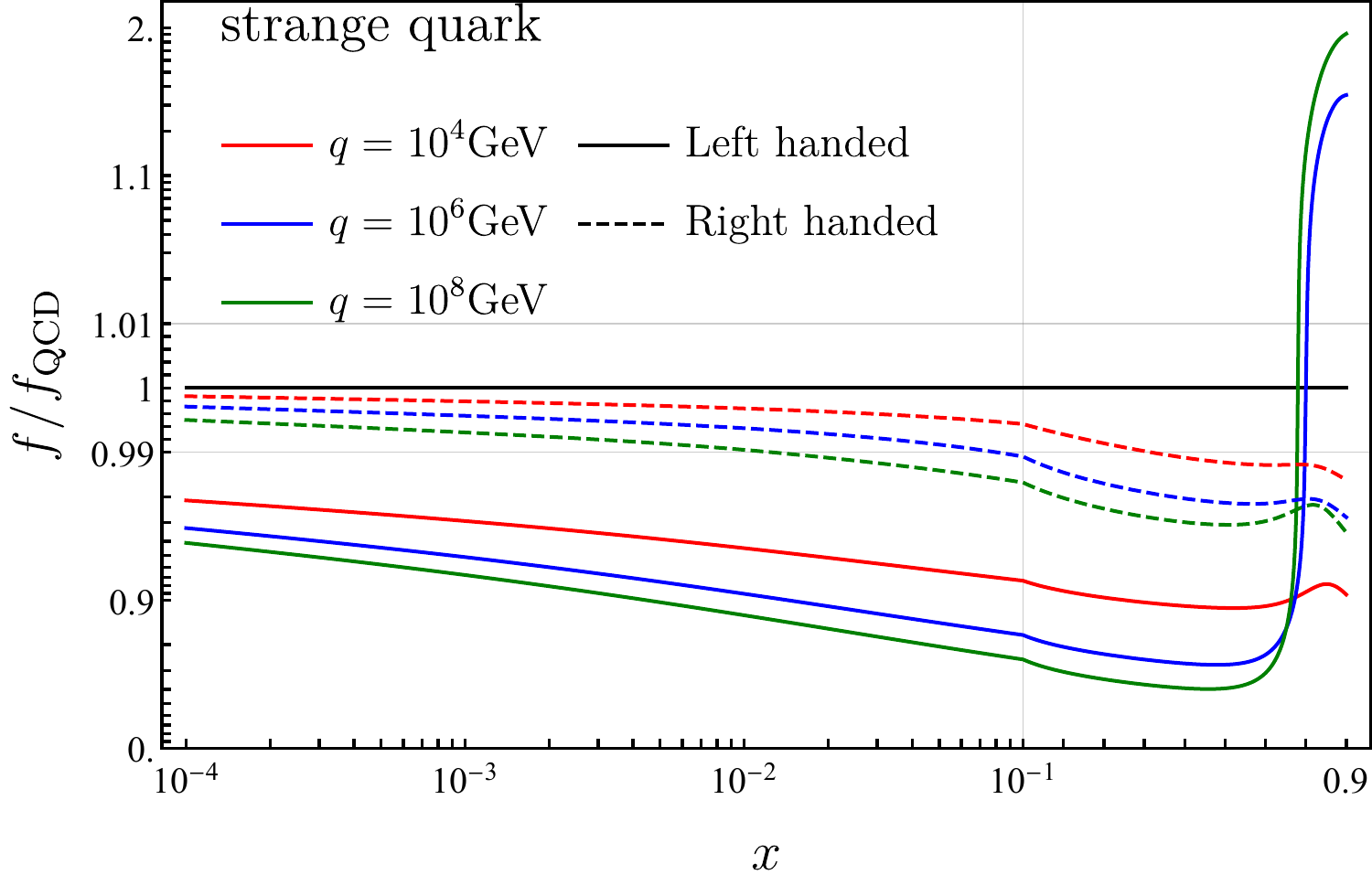}
  \includegraphics[scale=0.45]{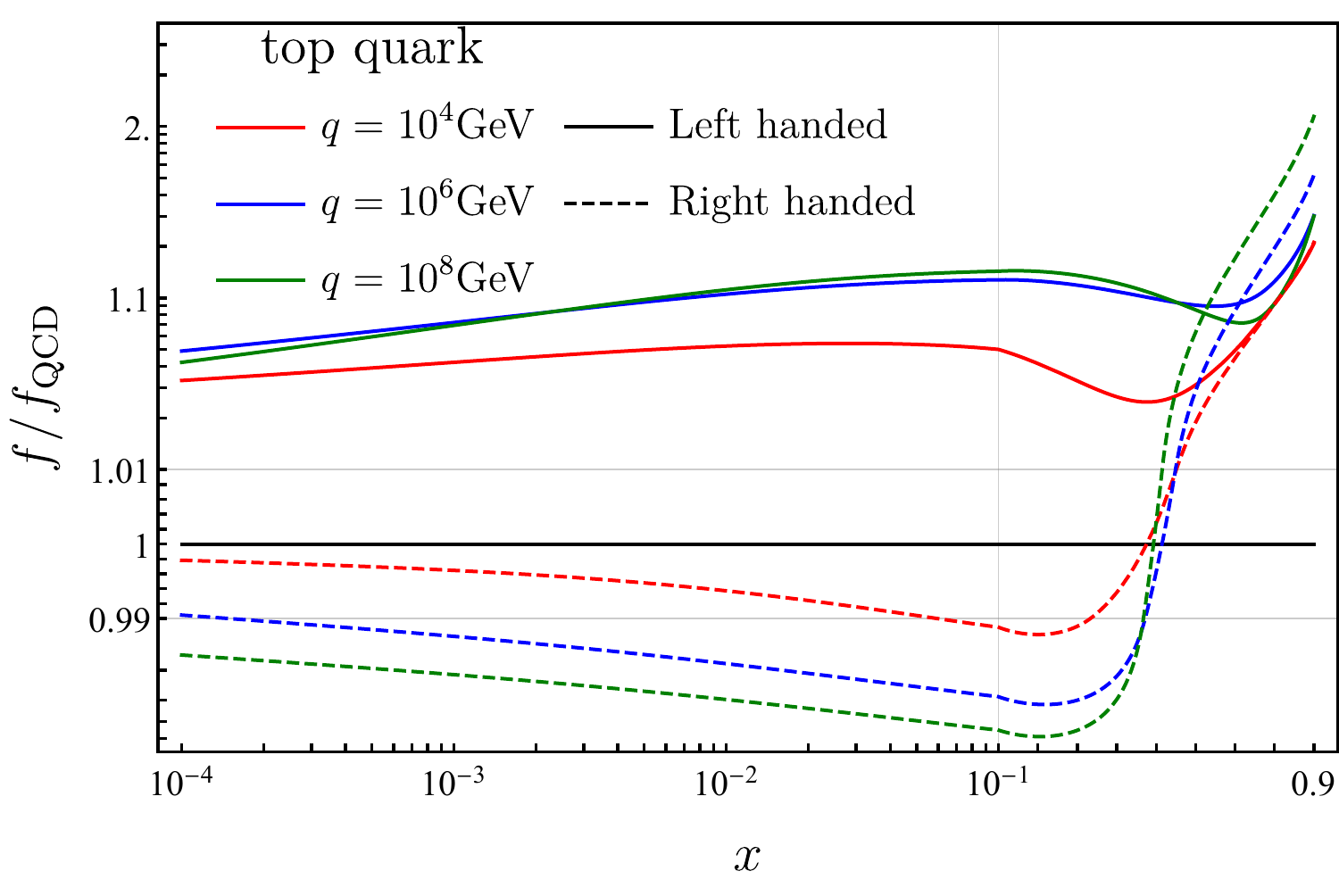}
  \includegraphics[scale=0.45]{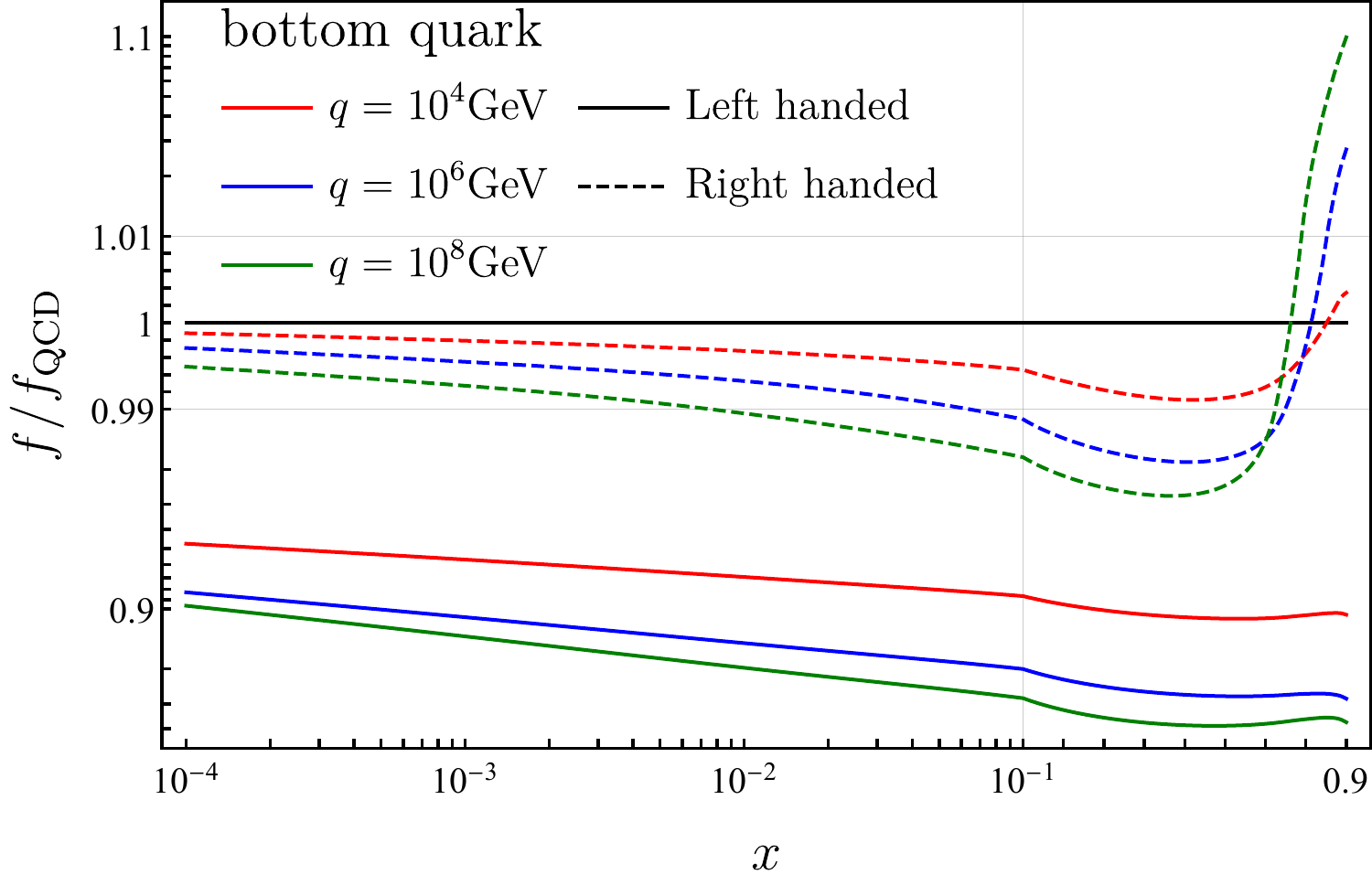}
  \includegraphics[scale=0.45]{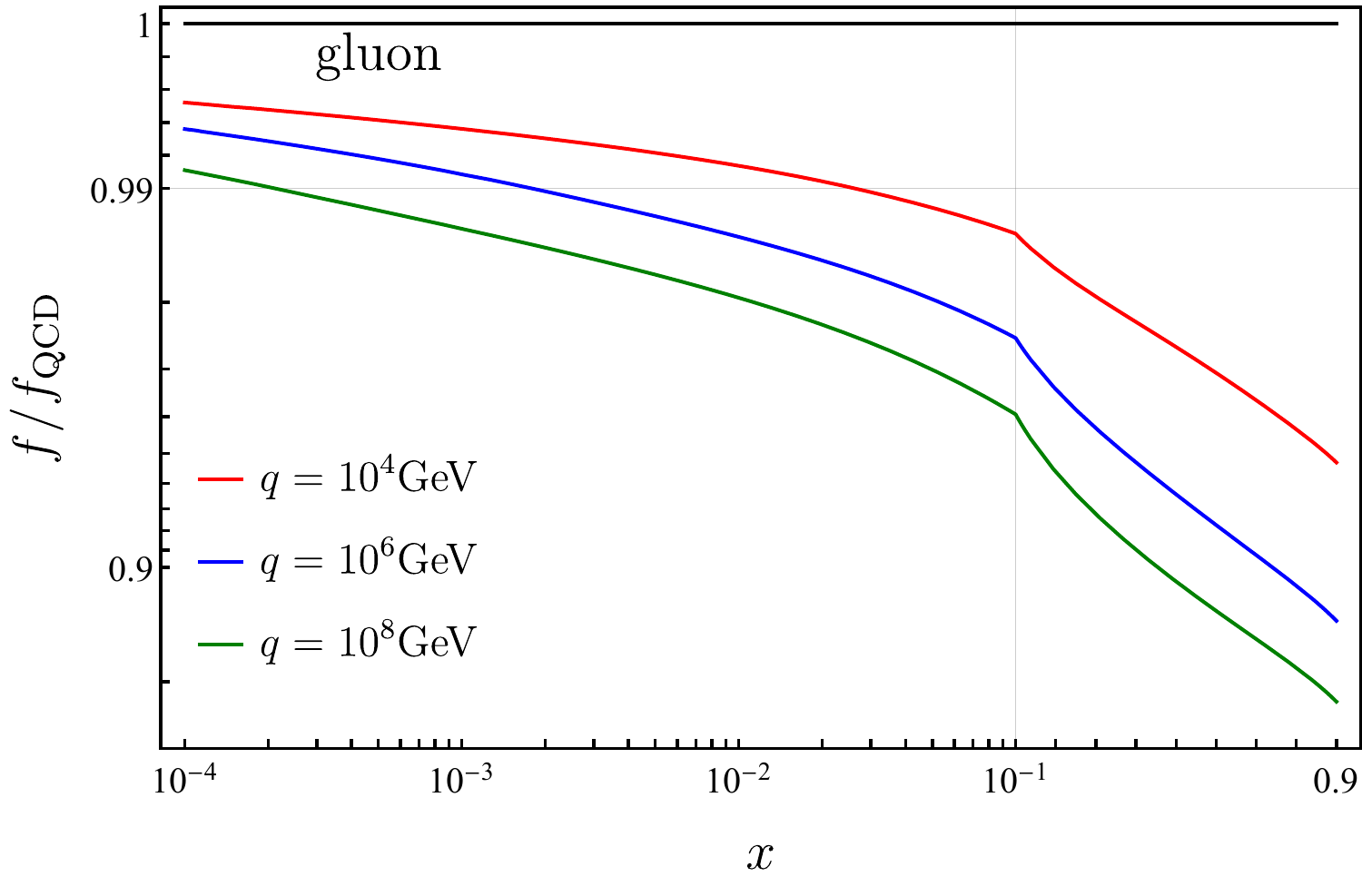}
   \caption{\label{fig:quarks}%
Quark and gluon PDFs in the full unbroken SM, divided by their values assuming
pure QCD evolution only.  Left- and right-handed quark chiralities are
solid and dashed, respectively. The thin gray lines show where the scales on the 
x- and/or y-axes switch between linear and logarithmic.}
}
The right-handed quark PDFs have no double-logarithmic component and
evolve to slightly lower values than pure QCD, due to energy loss through
the additional splitting $q_R \to q_R B$.

The asymmetry between left-handed charm and strange quarks also
evolves double-logarithmically towards zero, primarily through a more
rapid decrease of the strange PDF.  At high $x$ the behavior is more
complicated because the input CT14qed charm PDF is larger than the
strange above $x\sim 0.7$.  The right-handed quarks behave
qualitatively the same as those of the first generation. 

The left-handed top and bottom quarks also must evolve towards equal
values, which in this case means that the top has higher values that
in pure QCD, while the bottom evolution looks similar to strange,
relative to pure QCD.
The right-handed $b$-quark behaves qualitatively like the right-handed
quarks of the first and second generation, while the right-handed top
quark, being generated purely dynamically, behaves differently at
large $x$. Since the right-handed top has vanishing initial condition,
the splitting $t_R \to t_R B$, which would decrease the PDF, is
sub-dominant compared to the process $B \to t_R \bar t_R$. This means
that at large $x$ the right-handed top PDF is increased, rather than
decreased.
 
The effect on the gluon PDF is shown in last row of
Fig.~\ref{fig:quarks}.  While the effects are quite small up to
$q\sim 10^4$ GeV, at larger scales the back-reaction from the changing
quark PDFs is affecting the gluon PDF at an appreciable level.

\FIGURE{
	\centering
	\includegraphics[scale=0.45]{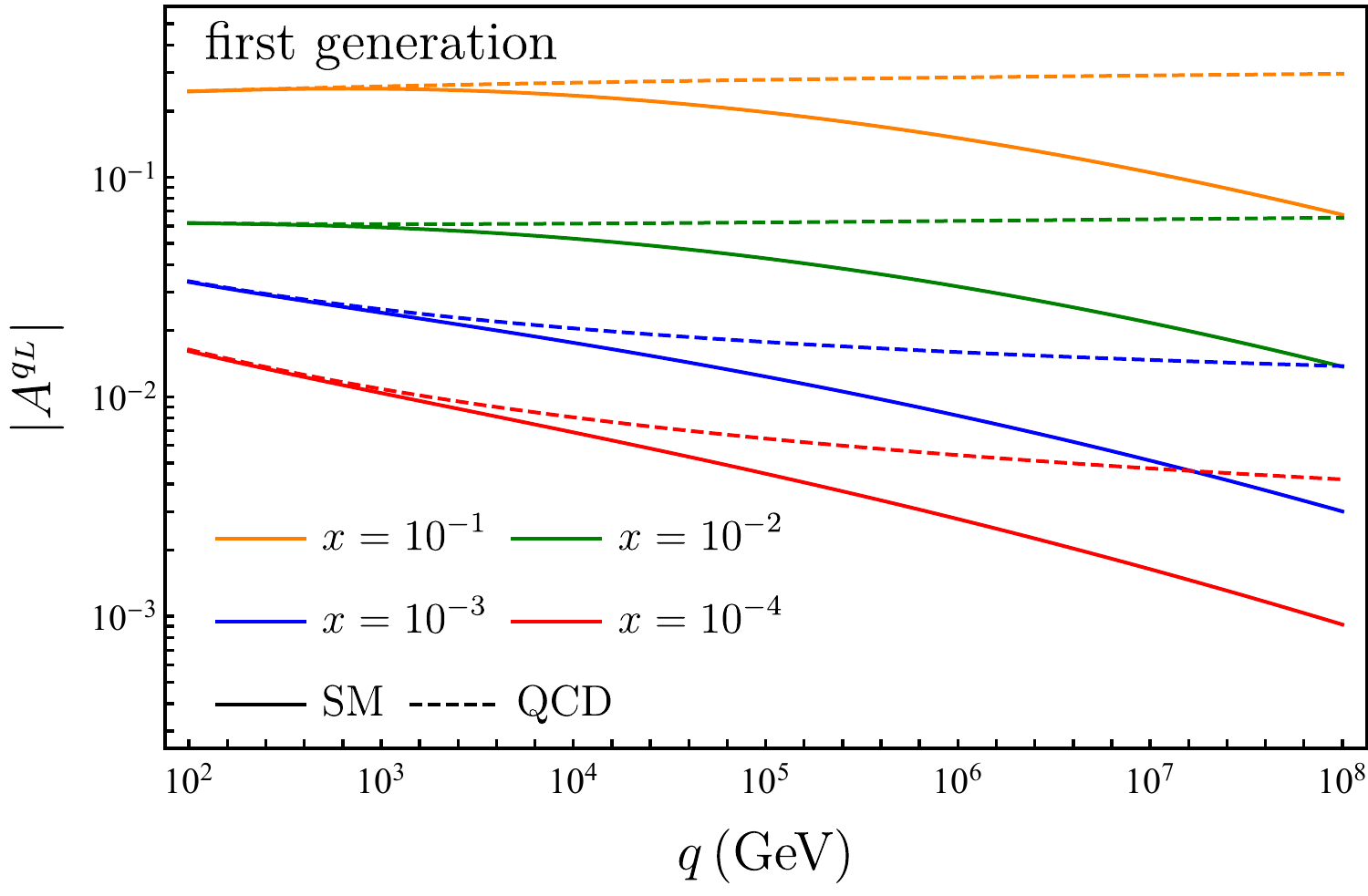}
	\includegraphics[scale=0.45]{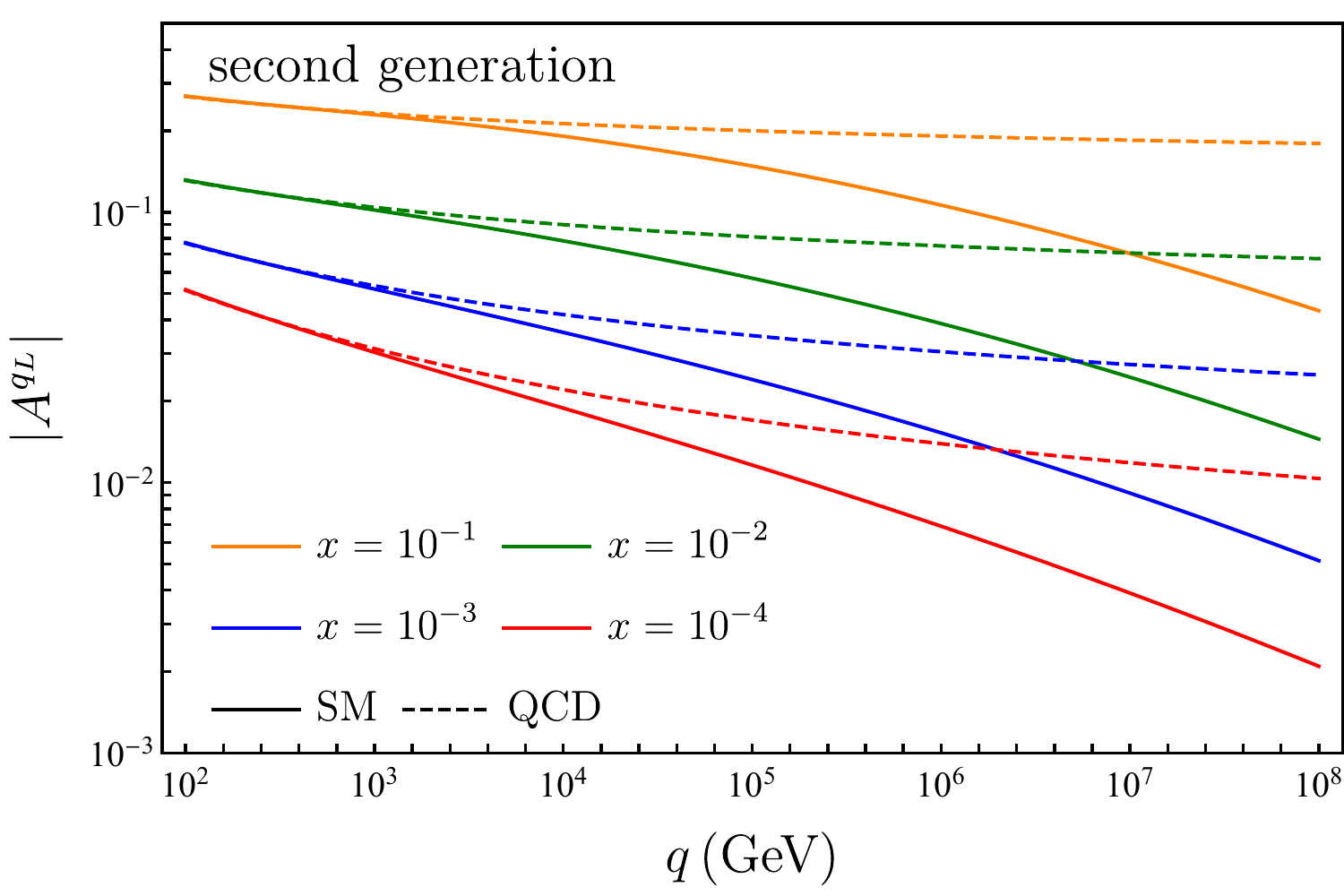}
	\includegraphics[scale=0.45]{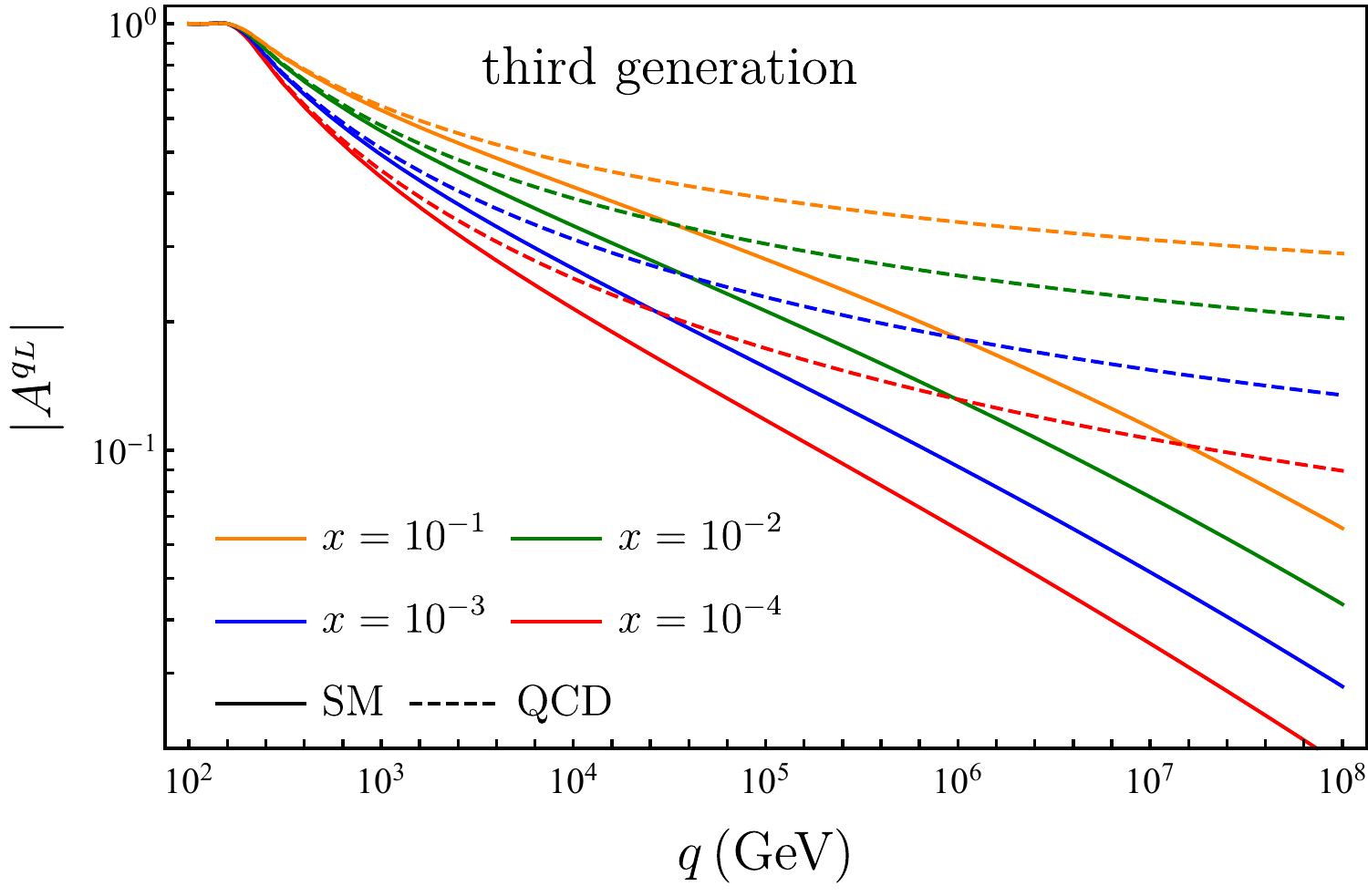}
	\caption{\label{fig:asymetry}%
		Asymmetry between up-isospin and down-isospin left-handed quark PDFs, defined in \eq{asymDef}, in the full unbroken SM, compared to the result when only QCD evolution is included.}
}

It is interesting to study how rapidly electroweak symmetry is
restored. To illustrate this, we show the asymmetry
\beq
\label{eq:asymDef}
A^{q_L} = \frac{f_{u_L} - f_{d_L}}{f_{u_L} + f_{d_L}},
\eeq
compared to the result if only QCD evolution were turned on. This
asymmetry ratio is shown in Figure~\ref{fig:asymetry} for the three
generation of quarks as a function of $q$, for various values of
$x$. For all generations the asymmetry decreases as $q$ gets larger,
driving the PDFs of the different isospin states towards each
other. The onset of the deviation from pure QCD is in the range $1-10$
TeV. The ratio between the full asymmetry and the result using only QCD
evolution is given by
\beq
A^{q_L} (x, q) \sim \left[\Delta_{f,2}(q)\right]^{4/3} A_{\rm QCD}^{q_L} (x, q)
\eeq
where $\Delta_{f,2}(q)$ is the fermion Sudakov factor, as given in \eq{Fdef}, independent
of the generation.

\FIGURE{
	\centering
	\includegraphics[scale=0.45]{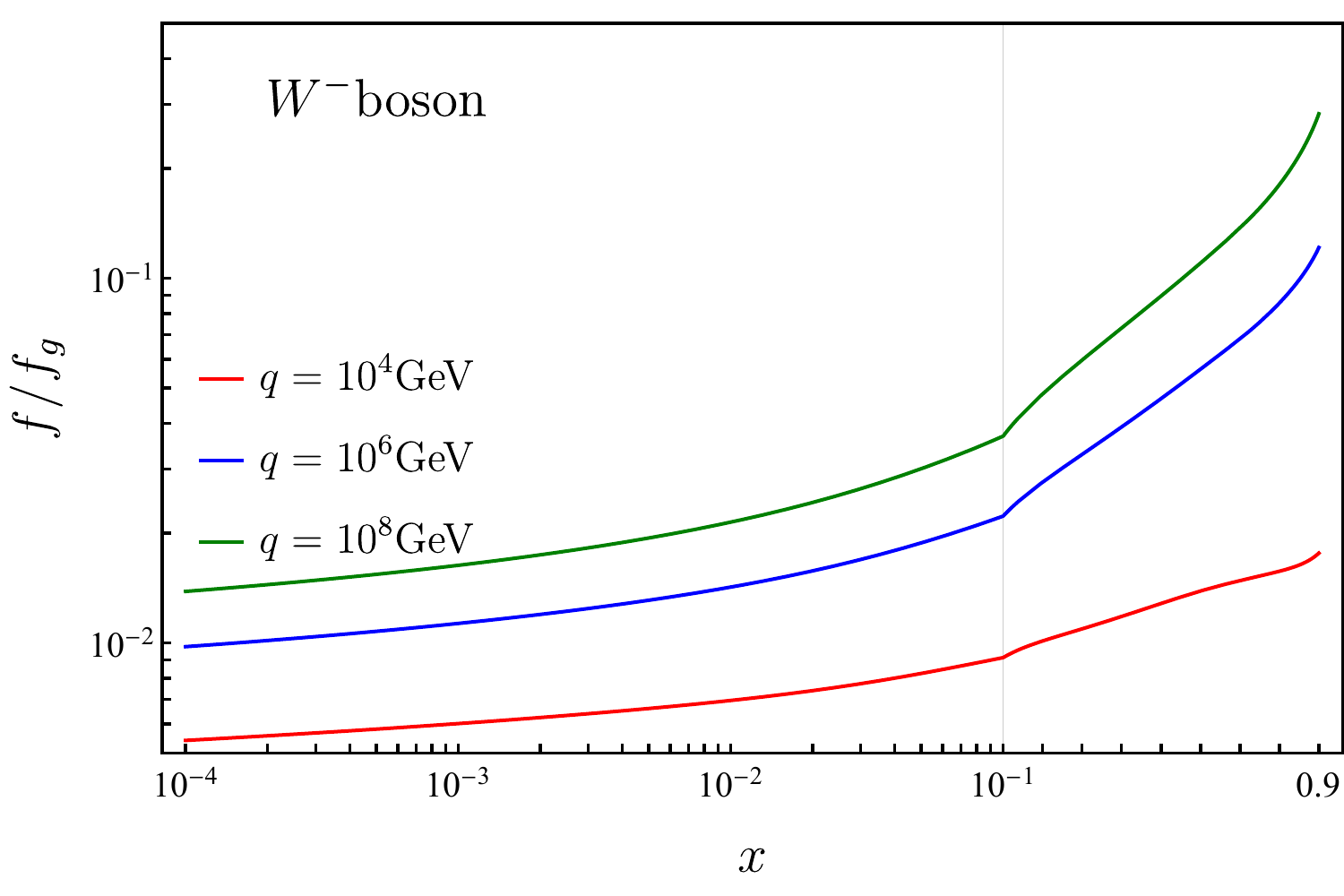}
	\includegraphics[scale=0.45]{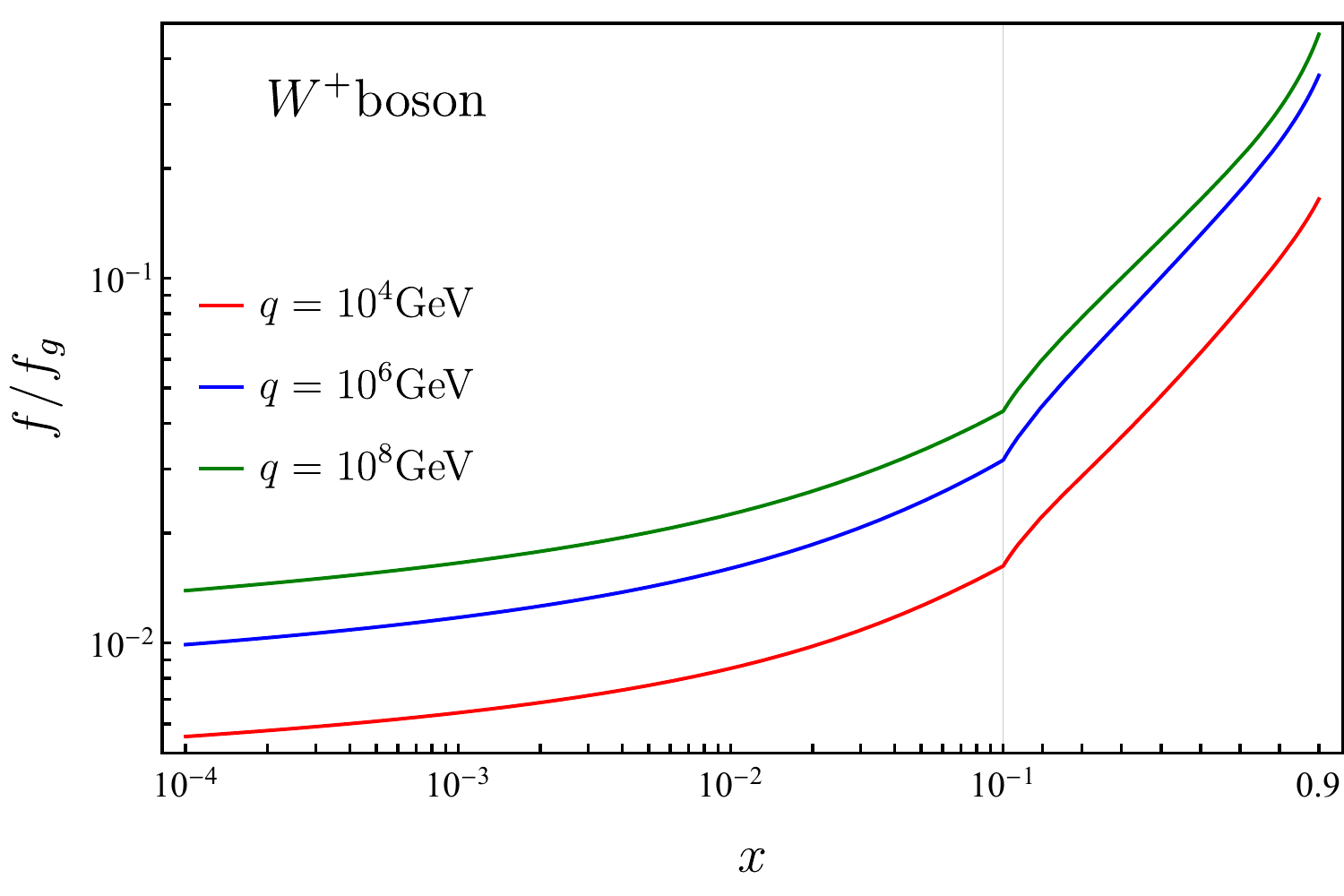}
	\includegraphics[scale=0.45]{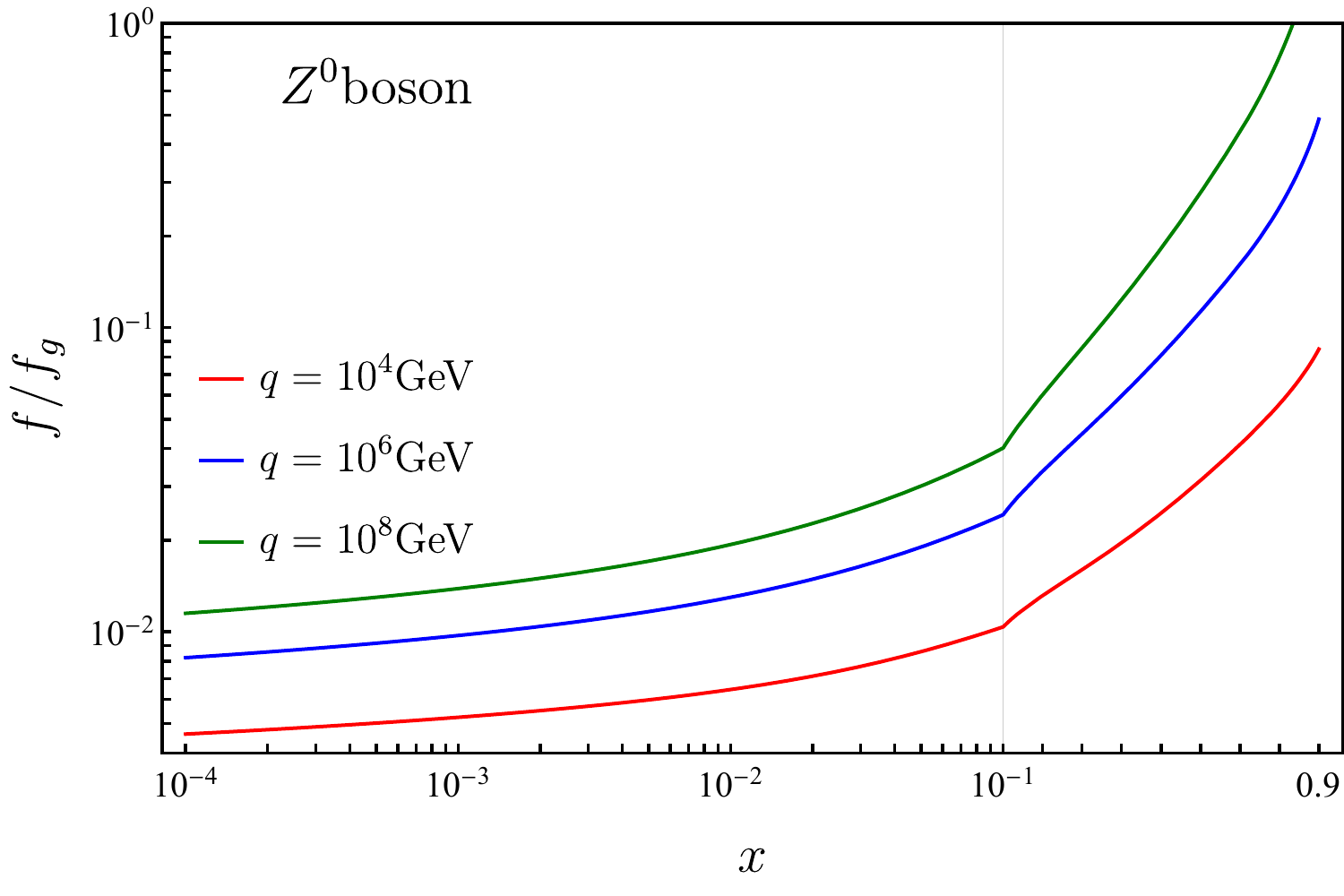}
	\includegraphics[scale=0.45]{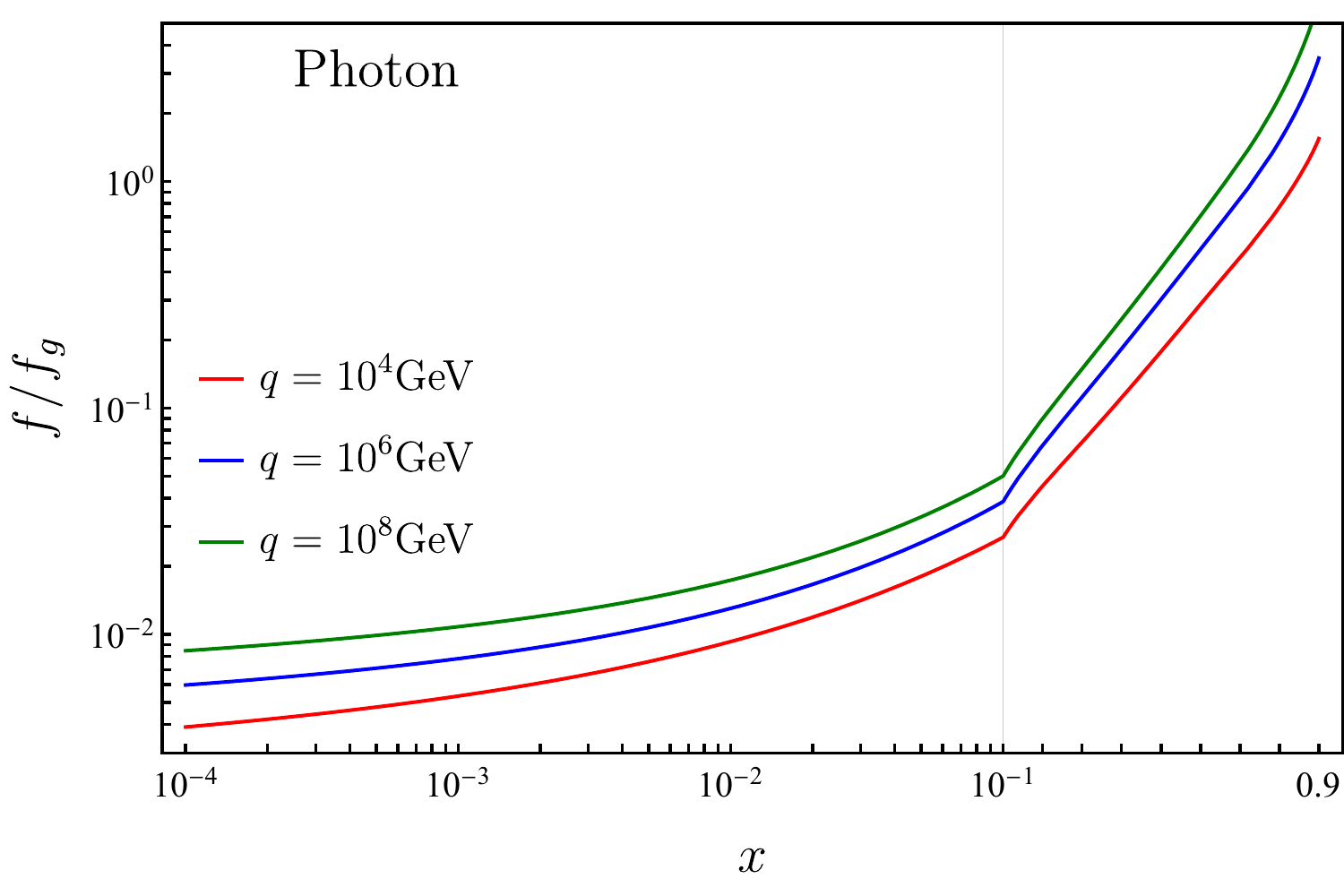}
	\includegraphics[scale=0.45]{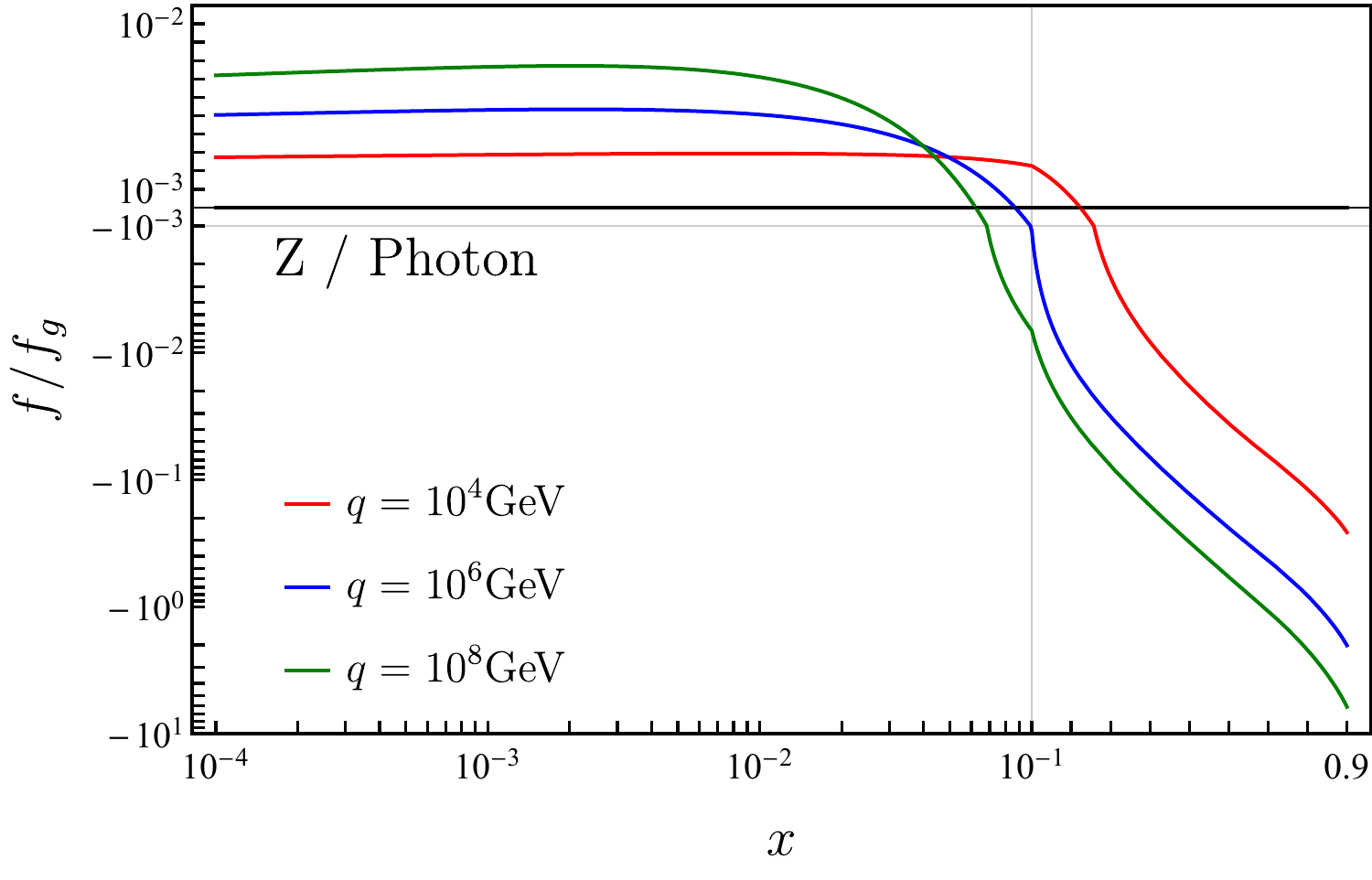}
	\caption{\label{fig:VectorBosons}%
		Electroweak bosons PDF normalized by the gluon PDF.  The thin gray lines show where the scales on the 
x- and/or y-axes switch between linear and logarithmic.
	}
}
\FIGURE{
	\centering
	\includegraphics[scale=0.45]{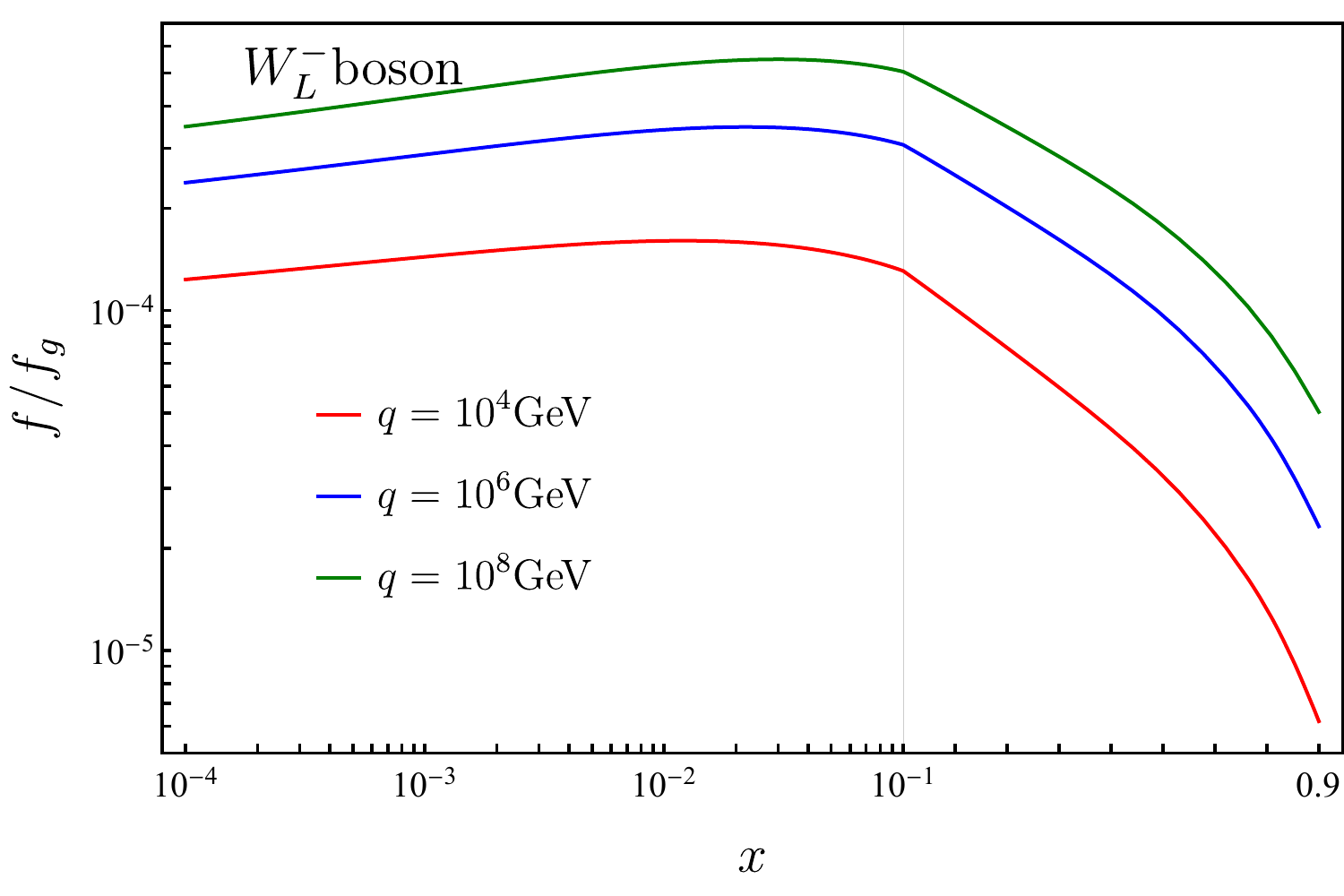}
	\includegraphics[scale=0.45]{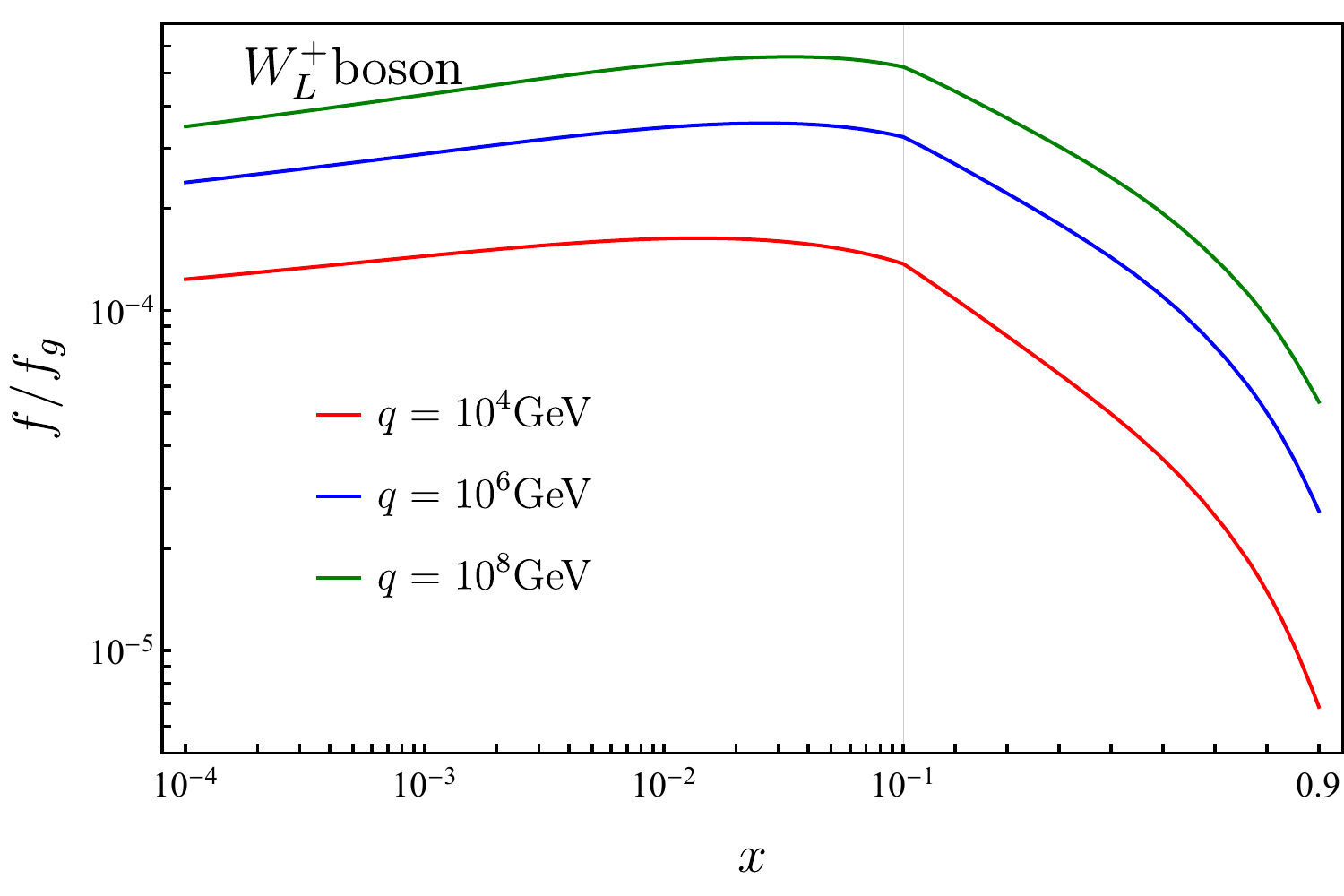}
	\includegraphics[scale=0.45]{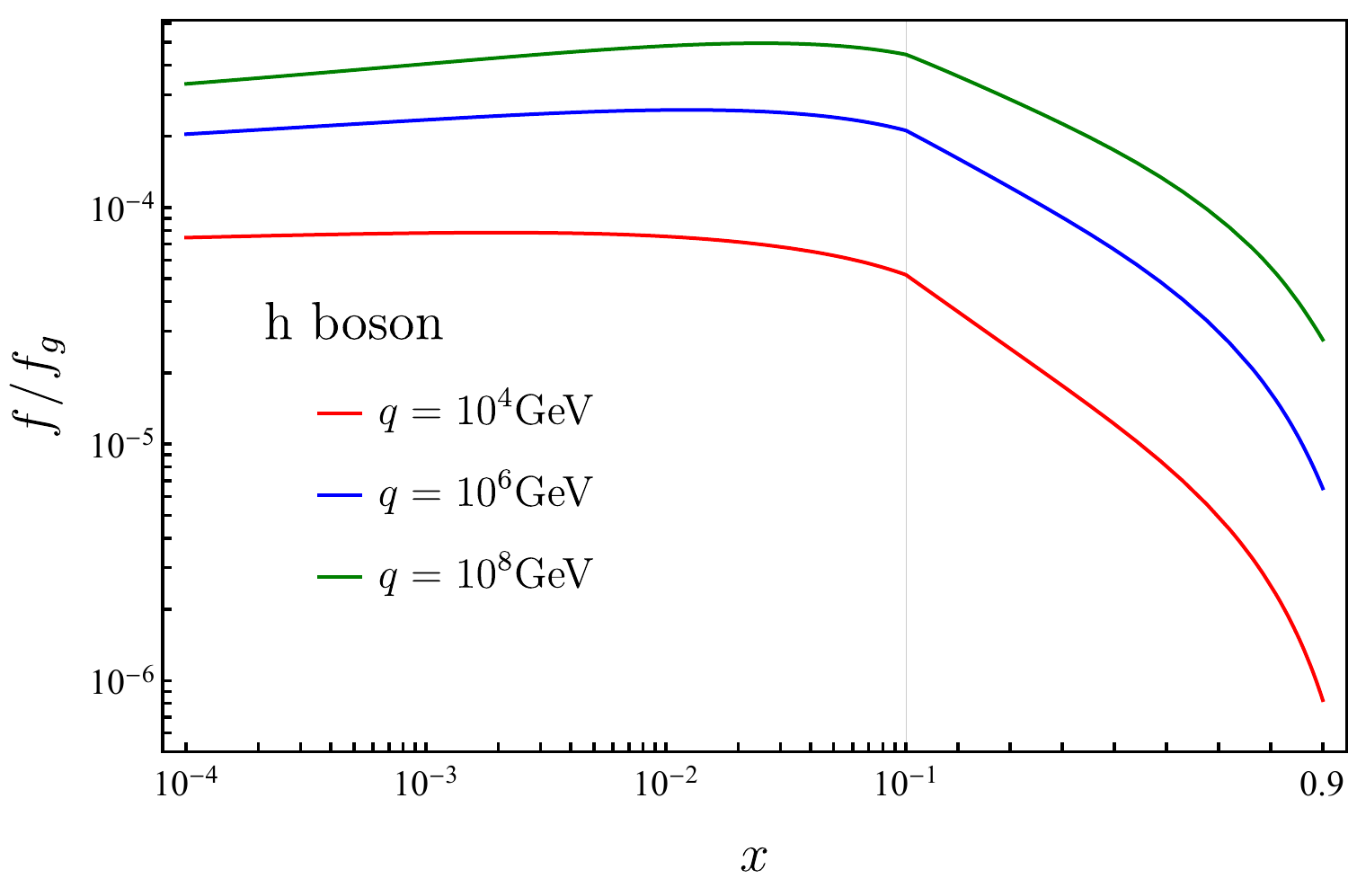}
	\includegraphics[scale=0.45]{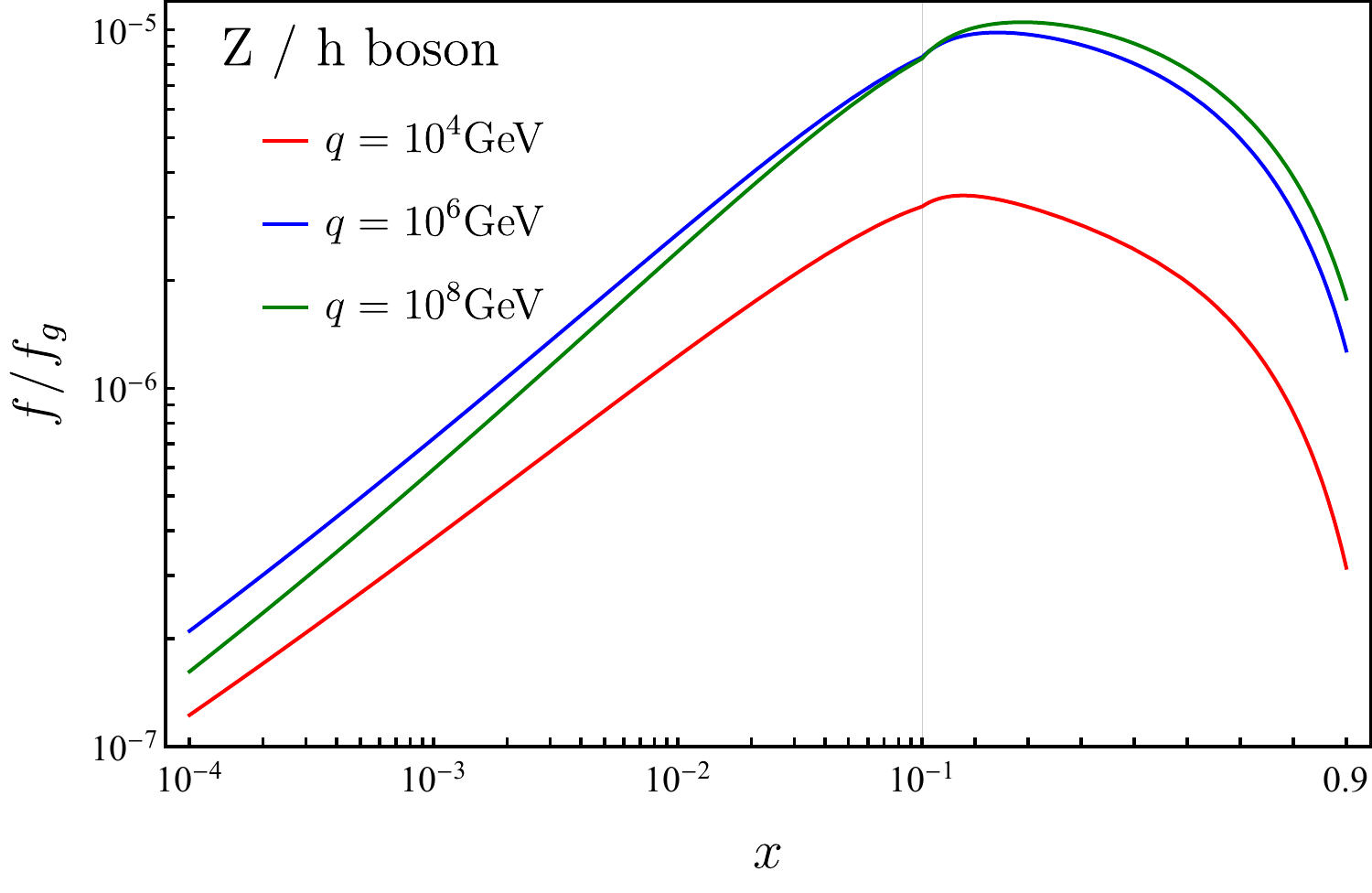}
	\caption{\label{fig:HiggsBosons}%
		 Longitudinal gauge and Higgs bosons PDFs normalized by the gluon PDF. The $Z_L$ PDF is the same as the $h$ PDF. The $hZ_L$ PDF is purely imaginary and we show the result divided by $i$. The thin gray line shows where the scales on the 
x- and/or y-axes switch between linear and logarithmic.
	}
}

\FIGURE{
	\centering
	\includegraphics[scale=0.45]{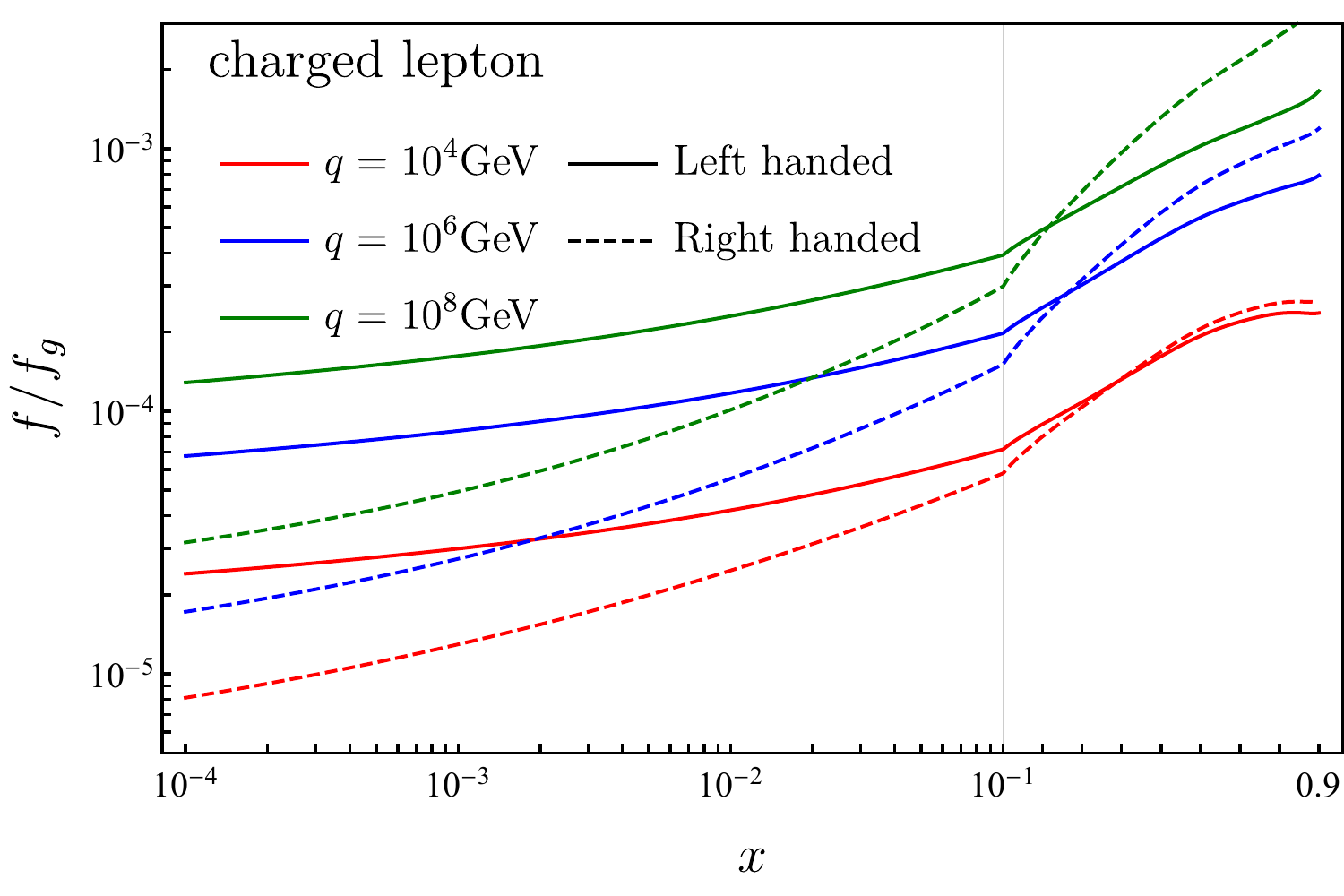}
	\includegraphics[scale=0.45]{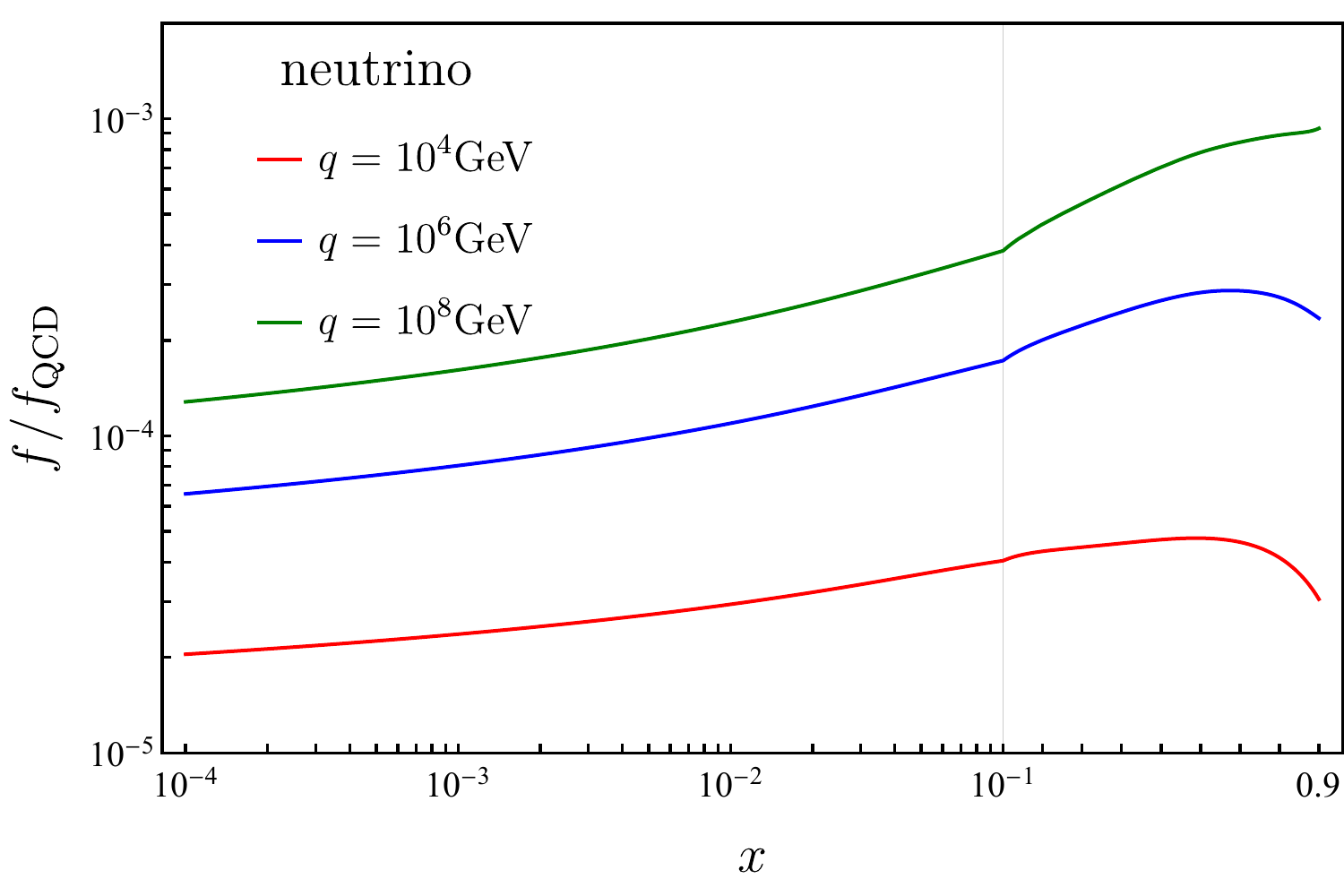}
	\caption{\label{fig:Leptons}%
		First generation lepton PDFs normalized by the gluon PDF. Since we treat leptons as massless, and all leptons have the same initial condition, the results for the other 2 generations are identical. The thin gray line shows where the scales on the 
x- and/or y-axes switch between linear and logarithmic.
	}
}

Next, we study the size of the PDFs of particles not charged under the strong interaction. Since these PDFs are only generated by emissions due to the U(1), SU(2) or Yukawa interactions, they are vanishing at all scales if one is including only  SU(3) evolution. The only exception is the photon, which has a non-vanishing initial condition at $q = 100$ GeV. Figure~\ref{fig:VectorBosons} shows results on the electroweak boson PDFs normalized to the gluon PDF, both evolved using the full Standard Model. One can see that the electroweak gauge boson PDFs become a significant fraction of the gluon PDF, especially at large values of $x$. 
The photon PDF is the largest mainly because it has a non-zero
input. The PDF for the $W^+$ boson is initially larger than the $W^-$
boson PDF at large $x$ because the $W^+$ is mainly generated through
emissions from the up-quark, whose PDF is larger than the down-quark
which mainly generates the $W^-$.  Since the difference between
$W^+$ and  $W^-$ has isospin 1, the $W^+$ evolves more slowly
and the $W^-$ more rapidly, so that they approach each other at
high $q$.  At low $x$ they are more similar as are the up-quark and
down-quark PDFs. The $Z^0$ PDF is similar to the $W^+$ but it is
smaller at low $x$ and larger at large $x$. The mixed $\gamma Z$ PDF is small and
positive at small $x$ and negative at large $x$. There is no constraint to
be positive definite for a mixed PDF as it is the product of two
amplitudes rather than the square modulus of one.  Its absolute value
becomes very large at large $x$ and $q$.

We also show the PDFs for the longitudinally polarized gauge bosons,
the Higgs boson, the mixed PDF between the Higgs and the $Z_L$ and the
leptons. The $Z_L$ PDF is the same as the Higgs in our approximation,
see \eq{fhphi3}, so we do not make a separate plot for it. The boson PDFs are shown in
Fig.~\ref{fig:HiggsBosons}, and the leptons in Fig.~\ref{fig:Leptons},
both normalized to the gluon. Both are expected to be much smaller
than the transverse vector boson PDFs, because they are generated via
a second order effect of emission from the vector bosons and
via Yukawa emission from the top and bottom quarks, which are
much smaller than the up and down quarks. The mixed PDF is even
smaller because it is generated by the asymmetry between transverse
$W^+$ and $W^-$ PDFs and the top and anti-top PDFs.  The
$W_L^+$ and $W_L^-$ PDFs are very similar, for the same reason.

\FIGURE{
	\centering
	\includegraphics[scale=0.45]{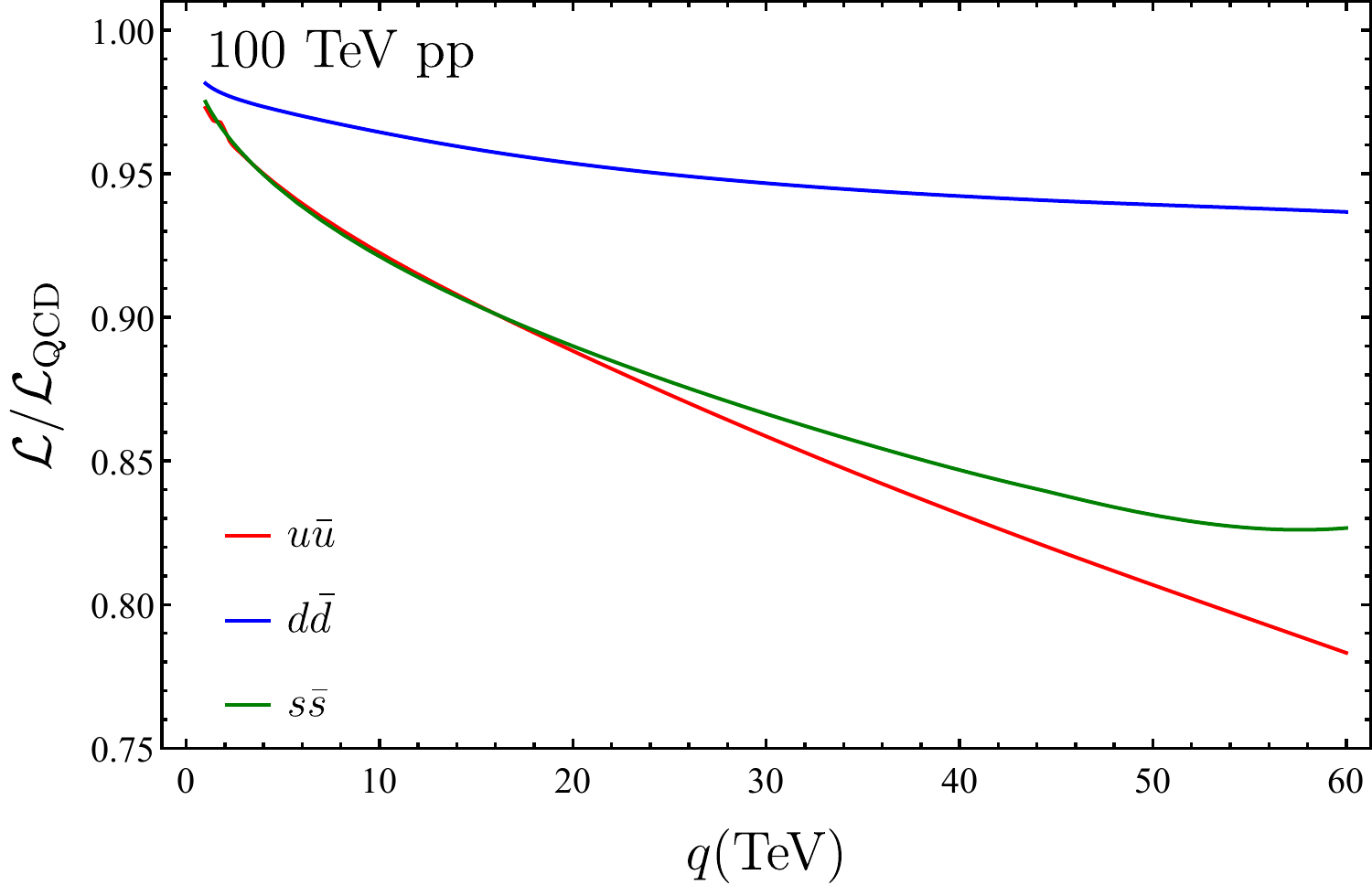}
	\includegraphics[scale=0.45]{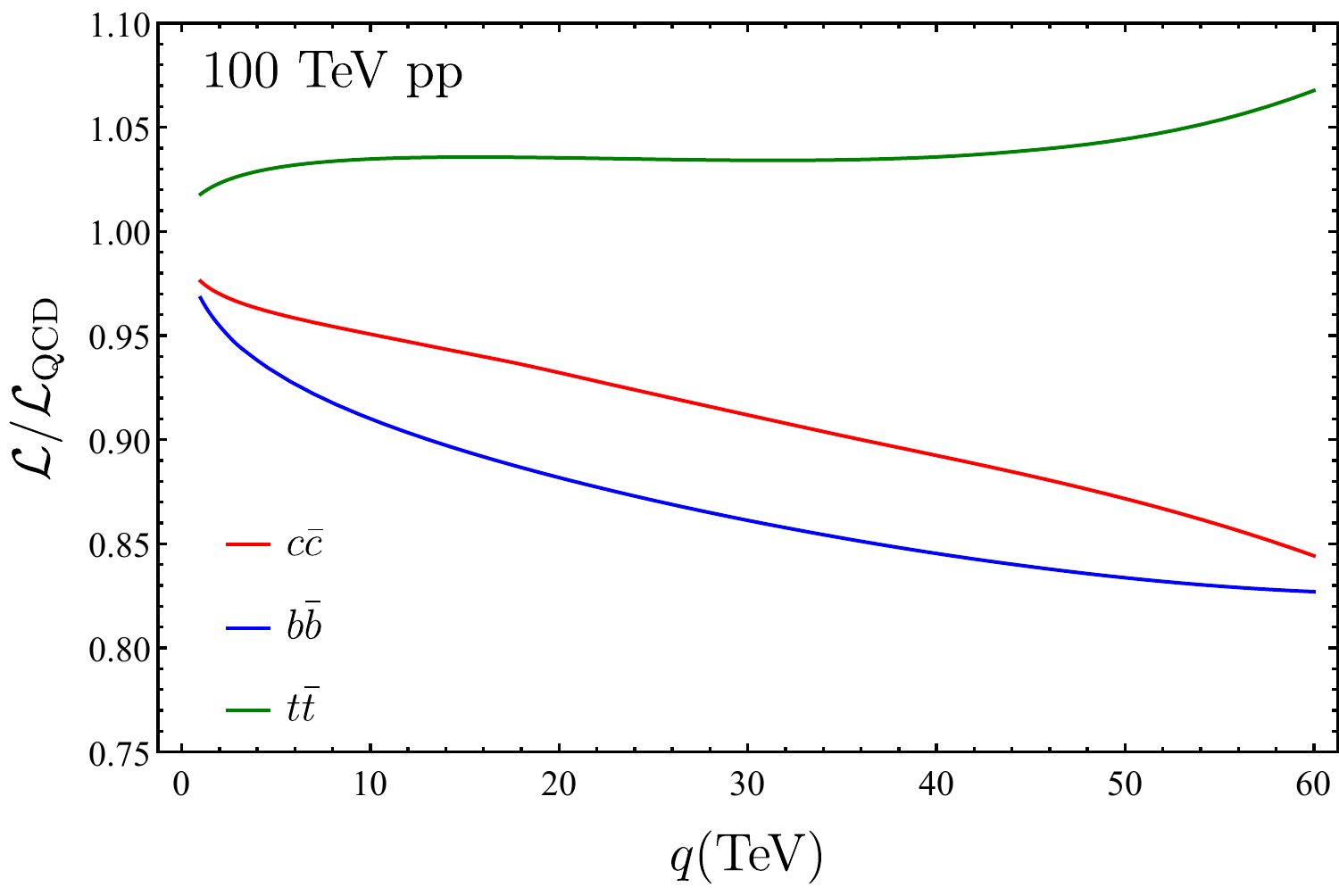}
	\caption{\label{fig:LumiQuark}%
Quark anti-quark luminosity in the full unbroken SM, divided by their values assuming pure QCD evolution only.}
}

\FIGURE{
	\centering
	\includegraphics[scale=0.45]{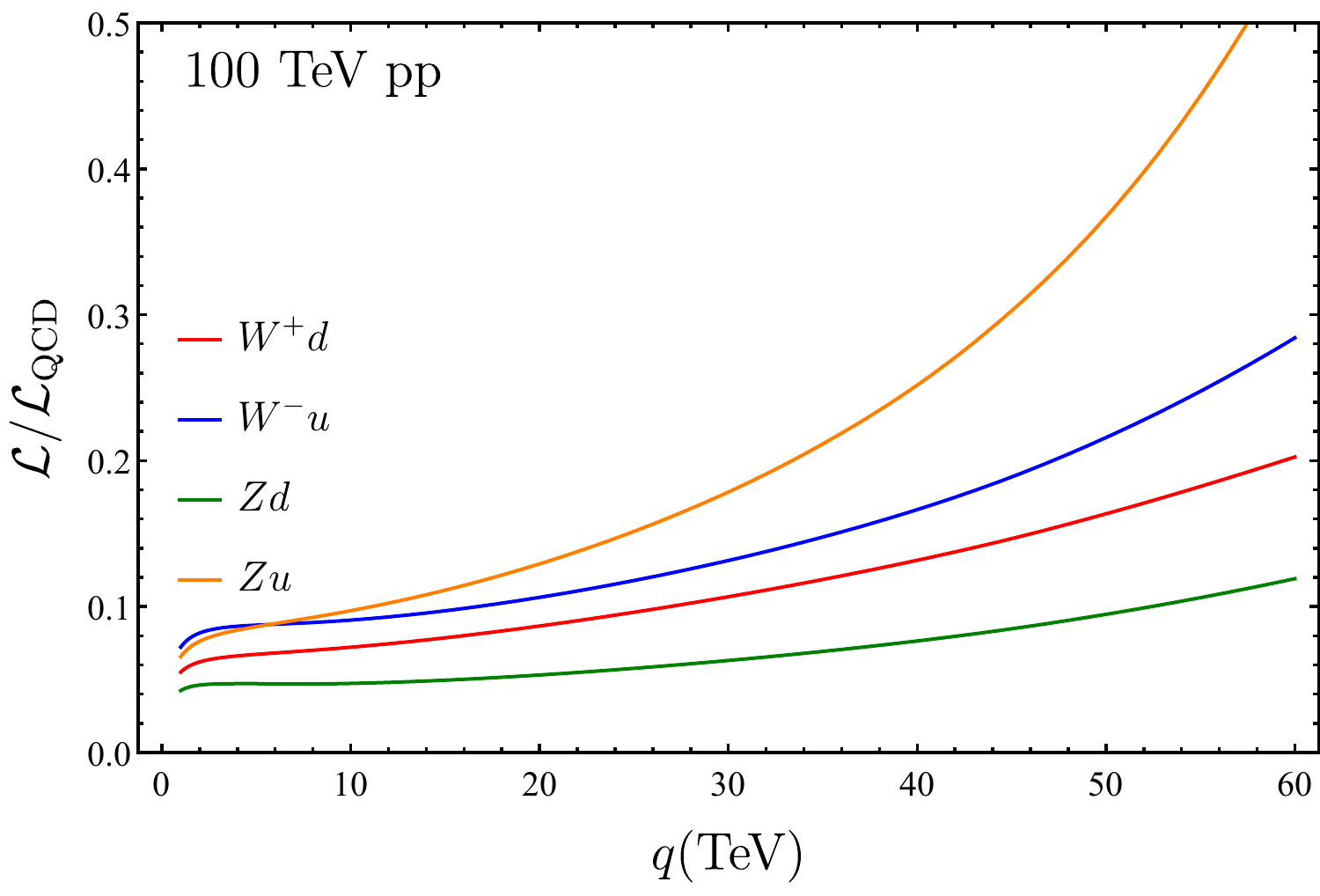}
	\includegraphics[scale=0.45]{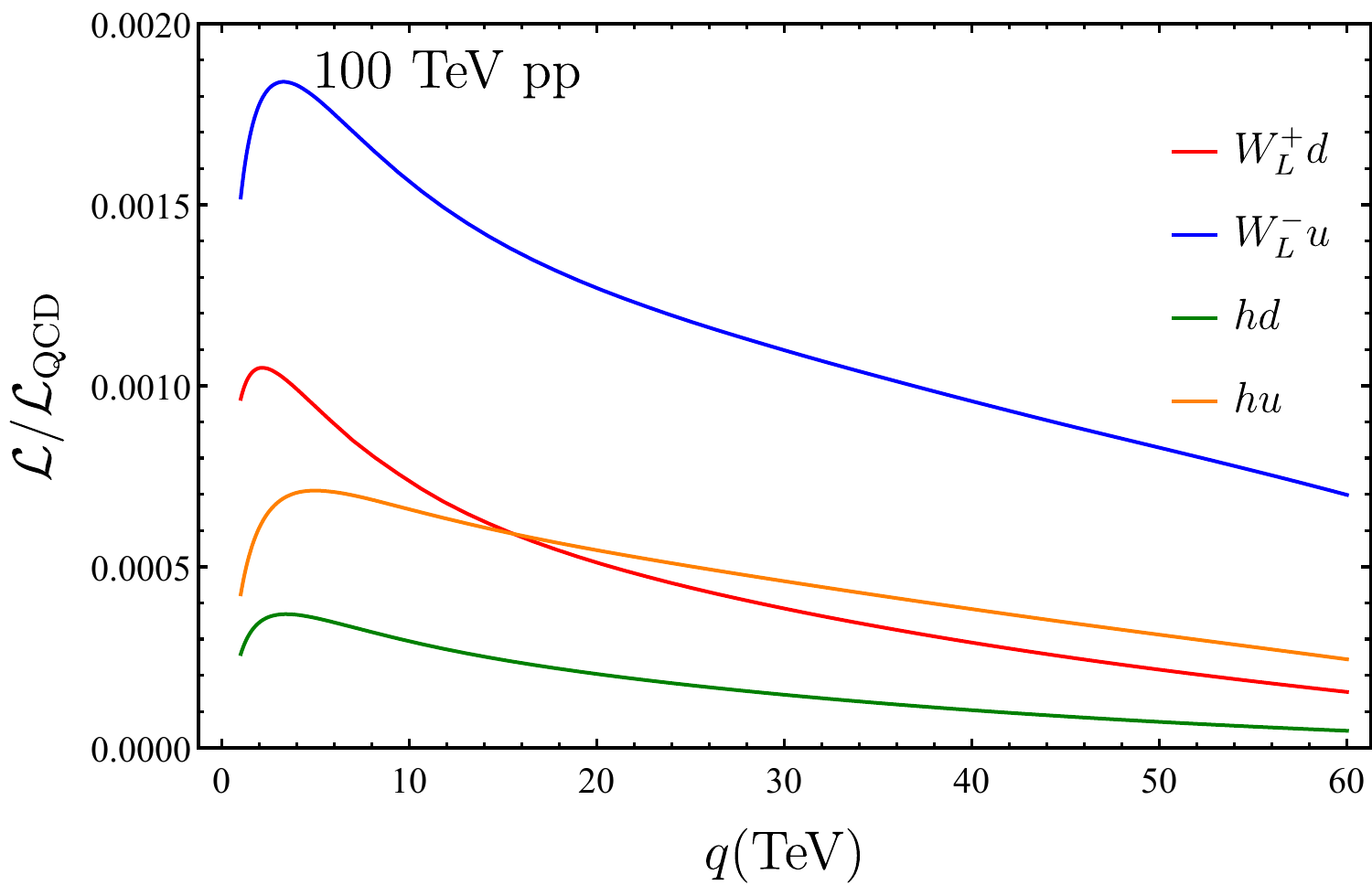}
	\caption{\label{fig:LumiVectorHiggs}%
$V q$ and $Hq$ luminosity in the full unbroken SM, divided by the average of $u \bar u$ and $d \bar d$ luminosity.}
}

As a final result, we study several parton luminosities, choosing a
future 100 TeV $pp$ collider as a reference. While the energy scales
that can be reached at such a collider are not quite large enough to
get ${\cal O}(1)$ effects, the effects of the full Standard Model
evolution are still numerically relevant. In
Figure~\ref{fig:LumiQuark} we show the $q_L \bar q_L$ luminosities
for the six different quark flavors, normalized to their values if only
QCD evolution is taken into account. One can see that all except the
$t\bar t$ luminosity are reduced appreciably from their
values if only QCD evolution were taken into account. This will affect
searches for $Z'$-like particles at a future 100 TeV collider. The $d
\bar d$ luminosity is decreasing more slowly as the double-logarithmic
evolution drives it larger than QCD at high $x$ (see Fig.~\ref{fig:quarks}).

We also show selected luminosities of vector bosons combined with
quarks, normalized to the average of the $u\bar u$ and $d\bar d$
luminosities. One can see that luminosities involving one transverse
vector boson become of comparable magnitude to the $q \bar q$
luminosities. Luminosities involving the longitudinal gauge and Higgs
bosons are much smaller. 

\section{Conclusions}
\label{sec:conc}
The energy regime around and beyond the electroweak scale is currently being explored by the LHC experiments, and so far they have found no firm evidence for physics beyond the Standard Model.  In the present paper, we have examined the consequences of assuming that the parton distributions of the proton continue to be described by the Standard Model up to very high energies, in the approximation that its symmetries are unbroken above the electroweak scale.

We have implemented numerically the full set of generalized DGLAP
evolution equations for all the parton species and interactions of the
unbroken SM in leading order.  
The input PDFs of 5 quark flavors, the gluon, photon and charged
leptons at a starting scale $q_0=100$ GeV for the full SM evolution are
obtained from parton and photon PDFs at 10 GeV by QCD plus QED
evolution.  The input left- and right-handed fermion PDFs are thus identical  at
scale $q_0$ but they evolve differently above that scale.  The top
quark PDFs (not present in the input) start to evolve from the top
mass scale.  The input photon is resolved into its U(1), SU(2) and
mixed components, which are evolved independently from scale $q_0$ and
reassembled into the photon and transversely polarized $Z^0$ at higher
scales.  The charged and longitudinal vector boson, Higgs and neutrino
PDFs are generated dynamically starting from zero at scale $q_0$.
This simplified treatment misses some symmetry-breaking effects around
the electroweak scale, but these are power-suppressed  at higher
scales and our results should provide a guide to the ways in which the
PDFs deviate from pure QCD evolution.

Amongst the most interesting features of the SM is the distinction between left- and right-handed fermions. The evolution of the right-handed PDFs deviates little from pure QCD, owing to the weakness of the U(1) interaction.  The left-handed PDFs generally deviate from pure QCD at the 5-10\% level by 10 TeV.

Another important SM characteristic is the restoration of isospin
symmetry at high scales.  This is manifest in the decreasing asymmetry
between the up- and down-type quark PDFs, which sets in at $1-10$ TeV,
the up-type being pulled down in the first generation and conversely
in the third.  The suppression of the asymmetry is a
double-logarithmic effect that can be treated in fixed order at
present energies but is resummed to all orders in the evolution.

The electroweak bosons are generated quite copiously, the $W^+$ in particular at high $x$ due to splitting of valence up quarks into $dW^+$.  The photon and $Z^0$ PDFs also grow rapidly, eventually exceeding the gluon at high $x$.  The PDFs of the longitudinal vector bosons, the Higgs boson and the leptons are generally much smaller as they arise from second-order splittings.

Finally, we have used the generated PDFs to present some parton-parton luminosities at a 100 TeV $pp$ collider. These results are just an illustration of the size of the effects that can be expected at such a future collider, and a more detailed phenomenological analysis will be presented in a forthcoming publication.

In conclusion, we find a rich structure in the proton when probed beyond the electroweak scale.  The associated PDFs are interesting and useful in their own right.  They also represent a key component of event generators that aim to embody the full Standard Model in initial-state parton showering, a topic we plan to explore further.

\acknowledgments
We thank Gavin Salam and Denis Comelli for comments on the manuscript.
This work was supported by the Director, Office of Science, Office of
High Energy Physics of the U.S. Department of Energy under the
Contract No. DE-AC02-05CH11231 (CWB, NF), and partially supported by
STFC consolidated grant ST/L000385/1 (BRW).

\appendix
\section{Equations used in the forward evolution}
\label{app:forward}

\subsection{SU(3) interaction}
\begin{itemize}
\item $\mathbf T = 0$ and ${\mathrm{CP}} = +$:
\beqn
\left[ q\frac{\pd}{\pd q} f^{0+}_q \right]_3  &=& \frac{\a_3}\pi \left[C_F P^+_{ff,G} \otimes f^{0+}_q
+T_R P^R_{fV,G}\otimes f_g\right],\\
 \left[ q\frac{\pd}{\pd q}f_g \right]_3  &=& \frac{\a_3}\pi\left[C_A P^+_{VV,G}\otimes f_g+ C_F
P^R_{Vf,G}\otimes f^{0+}_{\sum_g}\right]\,.
\eeqn
Here
\beq
f^{0+}_{\sum_g} = 4 \sum_{q_L} f^{0+}_{q_L}+ 2 \sum_{q_R} f^{0+}_{q_R}
\,,\eeq
where the sums run over all left-handed quark doublets and all right-handed quarks. The factors of $4$ and $2$ are due to the different normalizations in \eqs{fLIsospin}{fRIsospin}. 

\item All other states:
\beqn
\left[ q\frac{\pd}{\pd q} f_q \right]_3  &=& \frac{\a_3}\pi C_F P^+_{ff,G} \otimes f_q
\,.\eeqn
\end{itemize}

\subsection{U(1) interaction}

\begin{itemize}
\item $\mathbf T = 0$ and ${\mathrm{CP}} = +$:
\beqn
\left[ q\frac{\pd}{\pd q} f^{0+}_f \right]_1 &=& \frac{\a_1}\pi Y_i^2\left[P^+_{ff,G}\otimes f^{0+}_f
+N_f P^R_{fV,G}\otimes f_B\right],\\
\left[ q\frac{\pd}{\pd q} f_B \right]_1 &=& \frac{\a_1}\pi\left[P^V_{B,1} f_B +
P^R_{Vf,G}\otimes f^{0+}_{\sum_B f} + P^R_{VH,G} \otimes f^{0+}_H \right]\,,\\
\left[ q\frac{\pd}{\pd q} f^{0+}_H \right]_1 &=& \frac{\a_1}\pi \frac{1}{4} \left[
P^+_{HH,G}\otimes f^{0+}_H + P^R_{HV,G} \otimes f_B\right]\,,
\eeqn
where
\beq
f^{0+}_{\sum_B f} = 4 \sum_{f_L} Y_{f_L}^2 f^{0+}_{f_L}+ 2 \sum_{f_R} Y_{f_R}^2f^{0+}_{f_R}
\,.
\eeq

\item ${\mathbf T} = 1$ and ${\mathrm{CP}} = +$:
\beq
\left[ q\frac{\pd}{\pd q} f^{1+}_{BW} \right]_1 =\frac{\a_1}\pi \frac 12
P^V_{B,1} f^{1+}_{BW}
\,.\eeq

\item All other states:
\beqn
\left[ q\frac{\pd}{\pd q} f_f \right]_1 &=& \frac{\a_1}\pi Y_f^2P^+_{ff,G}\otimes f_f
,\\
\left[ q\frac{\pd}{\pd q} f_H \right]_1 &=& \frac{\a_1}\pi \frac{1}{4} 
P^+_{HH,G}\otimes f_H \,.
\eeqn

\end{itemize}

\subsection{SU(2) interaction}

\begin{itemize}
\item $\mathbf T = 0$ and ${\mathrm{CP}} = +$:
\beqn\label{eq:SU2f0plus}
\left[ q\frac{\pd}{\pd q} f^{0+}_{f_L} \right]_2 &=& \frac{\a_2}{\pi}\frac 34\left[
  P^+_{ff,G}\otimes f^{0+}_{f_L}+ N_f P^R_{fV,G}\otimes f^{0+}_W\right] \,,\\
\left[ q\frac{\pd}{\pd q} f^{0+}_W \right]_2 &=& \frac{\a_2}\pi\left[2 P^+_{VV,G}\otimes
  f^{0+}_W+\sum_{f_L} P^R_{Vf,G}\otimes f^{0+}_{f_L} + P^R_{VH,G}\otimes f^{0+}_H\right] \,,
  \\
\left[ q\frac{\pd}{\pd q} f^{0+}_H \right]_2 &=& \frac{\a_2}\pi
 \frac{3}{4}  \left[ P^+_{HH,G}\otimes f^{0+}_H + P^R_{HV,G} \otimes f^{0+}_W\right]\,.
\eeqn

\item $\mathbf T = 0$ and ${\mathrm{CP}} = -$:
\beqn
 \left[ q\frac{\pd}{\pd q}f^{0-}_{f_L} \right]_2 &=& \frac{\a_2}{\pi}\frac 34
  P^+_{ff,G}\otimes f^{0-}_{f_L}\,,\\
\left[ q\frac{\pd}{\pd q} f^{0-}_H \right]_2 &=& \frac{\a_2}\pi
 \frac{3}{4} P^+_{HH,G}\otimes f^{0-}_H\,.
\eeqn

\item $\mathbf T = 1$ and ${\mathrm{CP}} = +$:
\beqn
 \left[ \Delta_{f,2}^{4/3}q\frac{\pd}{\pd q}\frac{f^{1+}_{f_L}}{\Delta_{f,2}^{4/3}}\right]_2 &=& -\frac{\a_2}{\pi}\frac 14
  P^+_{ff,G}\otimes f^{1+}_{f_L} \\
\left[ \Delta_{H,2}^{4/3}q\frac{\pd}{\pd q} \frac{f^{1+}_H}{\Delta_{H,2}^{4/3}}\right]_2 &=& -
\frac{\a_2}\pi \frac{1}{4} P^+_{HH,G} \otimes f^{1+}_H\\
\left[\Delta_{V,2}^{1/2}q\frac{\pd}{\pd q}\frac{f^{1+}_{BW}}{\Delta_{V,2}^{1/2}}\right]_2 &=& 0\,.
\eeqn

\item $\mathbf T = 1$ and ${\mathrm{CP}} = -$:
\beqn
\left[ \Delta_{f,2}^{4/3}q\frac{\pd}{\pd q} \frac{f^{1-}_{f_L}}{\Delta_{f,2}^{4/3}} \right]_2 &=& \frac{\a_2}{\pi}\left[-\frac 14
  P^+_{ff,G}\otimes f^{1-}_{f_L}+\frac 12 N_f  P^R_{fV,G}\otimes
 f^{1-}_{W}\right] \\
\left[ \Delta_{V,2}^{1/2}q\frac{\pd}{\pd q}\frac{f^{1-}_{W}}{\Delta_{V,2}^{1/2}} \right]_2 &=& \frac{\a_2}\pi\left[P^+_{VV,G}\otimes  f^{1-}_{W}+\sum_{f_L} P_{Vf}\otimes
  f^{1-}_{f_L} + P_{VH}\otimes f^{1-}_{H}\right] \\
  \nn
\left[\Delta_{H,2}^{4/3}q\frac{\pd}{\pd q}\frac{f^{1-}_{H}}{\Delta_{H,2}^{4/3}}\right]_2 &=& \frac{\a_2}\pi\left[ -\frac{1}{4} P^+_{HH,G} \otimes f^{1-}_{H} + \frac{1}{2} \, P_{HV,G} \otimes f^{1-}_{W} \right]\,.
\eeqn

\item $\mathbf T = 2$ and ${\mathrm{CP}} = +$:
\beqn\label{eq:SU2f2plus}
\left[ \Delta_{V,2}^{3/2}q\frac{\pd}{\pd q} \frac{f^{2+}_{W}}{\Delta_{V,2}^{3/2}}\right]_2 &=& -\frac{\a_2}\pi P^+_{VV}\otimes
  f^{2+}_{W}\,.
\eeqn
\end{itemize}

\subsection{Yukawa interaction}
\begin{itemize}
\item $\mathbf T = 0$ and ${\mathrm{CP}} = +$:
\beqn
\left[q\frac{\pd}{\pd q}  f^{0+}_{q^3_L} \right]_Y &=&
\frac{\a_Y}{\pi}\biggl[P^V_{q^3_L,Y} f^{0+}_{q^3_L} + 
P^R_{ff,Y} \otimes f^{0+}_{t_R}  + N_c P_{fH,Y} \otimes f^{0+}_{H} \biggr]\\
\left[q\frac{\pd}{\pd q}  f^{0+}_{t_R} \right]_Y &=&
\frac{\a_Y}{\pi}\,2\,\biggl[P^V_{t_R,Y} f^{0+}_{t_R} + 
P^R_{ff,Y} \otimes f^{0+}_{q^3_L}  + N_C P_{fH,Y} \otimes f^{0+}_{H}\biggr]\\
\left[ q\frac{\pd}{\pd q} f^{0+}_{H} \right]_Y &=&
\frac{\a_Y}{\pi}\biggl[P^V_{H,Y} f^{0+}_{H} + 
P^R_{Hf,Y} \otimes f^{0+}_{\sum_H f}\biggr]
\,,\eeqn
where
\beq
f^{0+}_{\sum_H f} = f^{0+}_{t_R} + f^{0+}_{q_L^3} \,.
\eeq

\item $\mathbf T = 0$ and ${\mathrm{CP}} = -$:
\beqn
\left[q\frac{\pd}{\pd q}  f^{0-}_{q^3} \right]_Y &=&
\frac{\a_Y}{\pi}\biggl[P^V_{q^3_L,Y} f^{0-}_{q^3_L} + 
P^R_{ff,Y} \otimes f^{0-}_{t_R}  - N_c P_{fH,Y} \otimes f^{0-}_{H} \biggr]\\
\left[q\frac{\pd}{\pd q}  f^{0-}_{t_R} \right]_Y &=&
\frac{\a_Y}{\pi}\,2\,\biggl[P^V_{t_R,Y} f^{0-}_{t_R} + 
P^R_{ff,Y} \otimes f^{0-}_{q^3}  + N_C P_{fH,Y} \otimes f^{0-}_{H}\biggr]\\
\left[ q\frac{\pd}{\pd q} f^{0-}_{H} \right]_Y &=&
\frac{\a_Y}{\pi}\biggl[P^V_{H,Y} f^{0-}_{H} + 
P^R_{Hf,Y} \otimes f^{0-}_{\sum_H f}\biggr]
\,,\eeqn
where
\beq
f^{0-}_{\sum_H f} = f^{0-}_{t_R} - f^{0-}_{q_L^3} \,.
\eeq

\item $\mathbf T = 1$ and ${\mathrm{CP}} = +$:
\beqn
\left[ q\frac{\pd}{\pd q} f^{1+}_{q^3_L} \right]_Y &=&
\frac{\a_Y}{\pi}\biggl[P^V_{q^3_L,Y} f^{1+}_{q^3_L} - N_c P_{fH,Y} \otimes f^{1+}_{H}\biggr] \\
\left[ q\frac{\pd}{\pd q} f^{1+}_{H} \right]_Y &=&
\frac{\a_Y}{\pi}\biggl[P^V_{H,Y} f^{1+}_{H}  -
P^R_{Hf} \otimes f^{1+}_{q_L^3}\biggr]
\eeqn

\item $\mathbf T = 1$ and ${\mathrm{CP}} = -$:
\beqn
\left[ q\frac{\pd}{\pd q} f^{1-}_{t_L} \right]_Y &=&
\frac{\a_Y}{\pi}\biggl[P^V_{t_L,Y} f^{1-}_{t_L} + N_c
P_{fH,Y} \otimes f^{1-}_{H}\biggr] \\
\left[ q\frac{\pd}{\pd q} f^{1-}_{H} \right]_Y &=&
\frac{\a_Y}{\pi}\biggl[P^V_{H,Y} f^{1-}_{H} + 
P^R_{Hf,Y} \otimes f^{1-}_{q_L^3}\biggr]
\eeqn
\end{itemize}

\subsection{Mixed interaction}
(Our results here differ slightly from Ref.~\cite{Ciafaloni:2005fm}.)
\begin{itemize}
\item $\mathbf T = 1$ and ${\mathrm{CP}} = +$:
\beqn\label{eq:mixedevol2}
\left[ q\frac{\pd}{\pd q} f^{1+}_{f} \right]_M &=& \frac{\a_M}\pi \frac{Y_f}{2}N_f
P^R_{fV,G}\otimes f^{1+}_{BW}\,,\\
\left[q\frac{\pd}{\pd q} f^{1+}_{BW}\right]_M &=& \frac{\a_M}\pi
\left[
4\sum_{f_L} Y_f P^R_{Vf,G}\otimes f^{1+}_{f}  + 2 P^R_{VH,G} \otimes f^{1+}_{H}
\right]\,,\\
\left[ q\frac{\pd}{\pd q} f^{1+}_H \right]_M &=& \frac{\a_M}\pi \frac{1}{4} P^R_{HV,G} \otimes f^{1+}_{BW}\,,
\eeqn
\end{itemize}

\clearpage
\addcontentsline{toc}{section}{References}
\bibliographystyle{JHEP}
\bibliography{SMevol_paper}

\end{document}